\documentclass{jfm}
\usepackage{upmath,bm,amsmath,amssymb,natbib,graphicx,mathrsfs}
\usepackage{array,epic,eepic,color}

\newcommand{\ie}{i.e.\ }
\newcommand{\Rey}{\mbox{\textit{Re}}}% Reynolds number
\newlength{\piclength}

\title[The Boussinesq approximation in rapidly rotating flows]
      {The Boussinesq approximation in rapidly rotating flows}

\author[Jose M. Lopez, Francisco Marques and Marc Avila] {J\ls O\ls
  S\ls E\ns M.\ns L\ls O\ls P\ls E\ls Z,$^{1}$ F\ls R\ls A\ls N\ls
  C\ls I\ls S\ls C\ls O\ns M\ls A\ls R\ls Q\ls U\ls E\ls S$^1$ and
  M\ls A\ls R\ls C\ns A\ls V\ls I\ls L\ls A$^2$}

\affiliation{ $^1$Departament de F{\'\i}sica Aplicada,
  Univ. Polit\`ecnica de Catalunya, Barcelona 08034,
  Spain\\[\affilskip] $^2$Institute of Fluid Mechanics,
  Friedrich-Alexander-Universit\"at Erlangen-N\"urnberg, 91058
  Erlangen, Germany\\[\affilskip] }

\pubyear{} \volume{} \pagerange{}
\date{\today}
\begin{document}

\maketitle

\begin{abstract} 

In commonly used formulations of the Boussinesq approximation
centrifugal buoyancy effects related to differential rotation, as well
as strong vortices in the flow, are neglected. However, these may play
an important role in rapidly rotating flows, such as in astrophysical
and geophysical applications, and also in turbulent convection. We
here provide a straightforward approach resulting in a Boussinesq-type
approximation that consistently accounts for centrifugal effects. Its
application to the accretion-disk problem is discussed.  We
numerically compare the new approach to the typical one in fluid flows
confined between two differentially heated and rotating cylinders. The
results justify the need of using the proposed approximation in
rapidly rotating flows.

\end{abstract}

%%%%%%%%%%%%%%%%%%%%%%%%%%%%%%%%%%%%%%%%%%%%%%%%%%%%%%%%%%%%%%%%%%%%%%
\section{Introduction}

In $1903$ Boussinesq observed that: ``The variations of density can be
ignored except where they are multiplied by the acceleration of
gravity in the equation of motion for the vertical component of the
velocity vector'' \citep{Bou03}. This simple approximation has had a
far-reaching impact on many areas of fluid dynamics; it allows us to
approximate flows with small density variations as incompressible,
whilst retaining the leading order effects due to the density
variations. Moreover, it is of great importance both analytically and
numerically as it eliminates acoustic modes, which are challenging to
treat.  Many problems in fluid dynamics have been tackled with
Boussinesq-type approximations, rendering in most cases successful
results in good agreement with experiments. However, some problems
feature important physics neglected in the original Boussinesq
approximation. For example, in many investigations of systems subject
to rotation, the centrifugal term in the Navier-Stokes equations is
treated as a gradient and is absorbed into the pressure
\citep{Ch61}. Under this assumption centrifugal buoyancy enters the
hydrostatic balance but does not play a dynamic role, making an
analytical treatment of the equations possible. In contrast, the
inclusion of centrifugal terms in numerical simulations requires a
minimal coding and computing effort. Therefore, it should always be
included in the simulations \citep{RFRM06}, and whether it is
dynamically significant or not should be determined a posteriori.

In systems rotating at angular velocity $\Omega$ the dynamical role of
centrifugal buoyancy is straightforward to model. Typically, a term
acting in the radial direction and proportional to $\rho'\Omega^2$,
where $\rho'$ is the density variation, is added to the Navier-Stokes
equation \citep{BaPe67,HoHu69}. One example where this term has been
included is rotating Rayleigh-B\'enard convection. \citet{Har00}
studied the effect of centrifugal buoyancy using a self-similar and
perturbative approach, confirmed by numerical simulations in the
axisymmetric case \citep{BHL00}. More recently, \citet{MMBL07,LoMa09}
conducted full 3D simulations in the same geometry. All these
investigations show the relevance of centrifugal buoyancy in rotating
convection. In these studies the imposed temperature gradient is
parallel to gravity, while in the present work both gradients are
perpendicular, and additional centrifugal effects, besides the
traditional $\rho'\Omega^2$ term, are also included.

Note that in the traditional approach, described in the previous
paragraph, effects due to differential rotation or strong internal
vorticity, of especial importance in rapidly rotating flows, are
neglected. The increasing interest in these flows because of their
industrial (e.g. cyclonic dust collectors or vortex chambers) and
scientific (astrophysical and atmospheric turbulence) applications
\citep[see][]{ElRo98} motivates the development of a new
approximation, which we here undertake. It is based on the Boussinesq
approximation but it includes additional physical effects stemming
from the advection term in the Navier--Stokes equations. It allows it
to accurately cast rapidly rotating flows with mild variations of
density into an incompressible formulation. In section \S2, we
describe a systematic way to achieve this, and we provide two
different and easy to implement ways to account for centrifugal
buoyancy effects in rotating problems.

We compare the different ways of including centrifugal effects in the
Boussinesq-Navier-Stokes equations by numerically studying the linear
stability of fluid between two differentially rotating cylinders
subject to a negative radial temperature gradient. Apart from its
intrinsic interest, this setting has been widely used to model both
atmospheric \citep[][]{hide1965} and astrophysical flows
\citep[][]{PJS07}, where the fluid reaches high rotational speeds. Our
simulations show that the traditional Boussinesq approximation (\ie
with the $\rho'\Omega^2$ term) is valid in a wide range of
angular speeds. However, for rapidly rotating flows important
centrifugal effects arise. Here even the linear behaviour of the
problem is significantly different for both approximations, justifying
the application of our approximation to account for centrifugal
effects.

The paper is organised as follows. After introducing the new
approximation in \S2, we compare it in \S3 to other approximations used
in accretion disk models. Section \S4 gives a detailed description of
the system as well as the governing equations of the problem and its
linearisation. A brief description of the base flow is also
provided. Section \S5 introduces the Petrov-Galerkin method
implemented to discretize the equations. In \S6 the linear stability
of the system considering both ways to introduce the centrifugal
buoyancy is compared. Various cases of interest are analysed. In \S6.1
we consider fluid rotating as a solid body, whereas in \S6.2 shear is
introduced in the system. We study quasi-Keplerian rotation in \S6.2.1
and a system rotating close to solid body subjected to weak shear in
\S6.2.2. Discussion and concluding remarks are given in \S7.

%%%%%%%%%%%%%%%%%%%%%%%%%%%%%%%%%%%%%%%%%%%%%%%%%%%%%%%%%%%%%%%%%%%%%%

\section{Boussinesq-type approximation for the centrifugal term}

In rotating thermal convection or stratified fluids the
Navier-Stokes-Boussinesq equations are usually formulated in the
rotating reference frame, with angular velocity vector
$\boldsymbol{\Omega}$. The momentum equation in this non-inertial
reference frame includes four inertial body force terms \citep{Bat67},
also called d'Alambert forces:
\begin{equation}\label{NS1}
\begin{aligned}
  \rho(\partial_t & +\mathbf{u}\cdot\nabla)\mathbf{u}= -\nabla p
    +\nabla\cdot\boldsymbol{\sigma}+\rho\,\mathbf{f}-\rho\nabla \Phi \\
  & -\rho\mathbf{A}-\rho\,\boldsymbol{\alpha}\times\mathbf{r}
    -2\rho\,\boldsymbol{\Omega}\times\mathbf{u}
    -\rho\,\boldsymbol{\Omega}\times(\boldsymbol{\Omega}\times\mathbf{r}).
\end{aligned}
\end{equation}
Here $-\rho\mathbf{A}$ is the translation force due to the
acceleration $\mathbf{A}$ of the origin of the rotating reference
frame, $-\rho\,\boldsymbol{\alpha}\times\mathbf{r}$ is the azimuthal
force (also called Euler force) due to the angular acceleration
$\boldsymbol{\alpha}=d\boldsymbol{\Omega}/dt$,
$-2\rho\mathbf{u}\times\boldsymbol{\Omega}$ is the Coriolis force and
$-\rho\,\boldsymbol{\Omega}\times(\boldsymbol{\Omega}\times\mathbf{r})$
is the centrifugal force (all of them per unit volume).  In
\eqref{NS1}, $\rho$, $p$ and $\mathbf{u}$ are the density, pressure
and velocity field of the fluid, $\mathbf{r}$ is the position vector
of the fluid parcel, and $\Phi$ is the gravitational potential, so
$-\rho\nabla \Phi$ is the gravitational force. The term
$\rho\mathbf{f}$ accounts for additional body forces that may act on
the fluid. For a Newtonian fluid the stress tensor
$\boldsymbol{\sigma}$ reads
\begin{equation}
  \boldsymbol{\sigma}=-p\,\mathbb{I}+\mu(\nabla\mathbf{u}+
    \nabla\mathbf{u}^T)+\lambda\nabla\cdot\mathbf{u}\,\mathbb{I},
\end{equation}
where $\mathbb{I}$ is the identity tensor, $\mu$ is the dynamic
viscosity, and $\lambda$ is the second viscosity.

\subsection{The Boussinesq approximation in a rotating reference
  frame}\label{sec:Bou_rf}

In the Boussinesq approximation all fluid properties are treated as
constant, except for the density, whose variations are considered only
in the ``relevant'' terms. Density variations are assumed to be small:
$\rho=\rho_0+\rho'$, with $\rho_0$ constant and $\rho'/\rho_0\ll1$;
the $\rho'$ term usually includes the temperature dependence, density
variations due to fluid density stratification, density variations in
a binary fluid with miscible species of different densities, etc. With
this assumption the continuity equation reduces to
$\nabla\cdot\mathbf{u}=0$ and the fluid can be treated as
incompressible. As a direct consequence the shear stress term in the
momentum equation \eqref{NS1} simplifies to the vector Laplacian, \ie
$\nabla\cdot\boldsymbol{\sigma}=\mu\nabla^2\mathbf{u}$.

Identifying the relevant terms in the momentum equation is a more
delicate issue.  Any term in \eqref{NS1} with a factor $\rho$ splits
into two terms, one with a factor $\rho_0$ and the other with a factor
$\rho'$. If a $\rho_0$ term is not a gradient, it is the leading-order
term, and the associated $\rho'$ term may be neglected. If the
$\rho_0$ term is a gradient, it can be absorbed into the pressure
gradient and does not play any dynamical role, and therefore the
associated $\rho'$ term must be retained in order to account for the
associated force at leading order. This is exactly what happens with
the gravitational term: $-\rho_0\nabla \Phi=\nabla(-\rho_0\Phi)$,
which is absorbed into the pressure gradient term and we must retain
the $-\rho'\nabla \Phi$ term to account for gravitational buoyancy.
The same treatment must be applied to the translation and centrifugal
terms, yielding the gradient terms
\begin{equation}\label{gradients}
  -\rho_0\mathbf{A}-\rho_0\boldsymbol{\Omega}\times(\boldsymbol{\Omega}
  \times\mathbf{r})=\nabla\big(\,\frac{1}{2}\rho_0|\boldsymbol{\Omega}\times
  \mathbf{r}|^2 -\rho_0\mathbf{A}\cdot\mathbf{r}\big),
\end{equation}
as well as $-\rho'\mathbf{A}$ and
$-\rho'\boldsymbol{\Omega}\times(\boldsymbol{\Omega}\times\mathbf{r})$,
which must be also retained.

The $\rho_0$ part of the remaining terms in equation \eqref{NS1} (so
far, we have considered the gravitational, centrifugal and
translational forces) are not gradients, so they are retained as
leading order terms and the corresponding $\rho'$ terms are neglected,
leading to the Boussinesq approximation equations in the rotating
reference frame:
\begin{equation}\label{Bou_rot_frame}
\begin{aligned}
  \rho_0 & (\partial_t+\mathbf{u}\cdot\nabla)\mathbf{u}= -\nabla p^*
    +\mu\nabla^2\mathbf{u}+\rho\,\mathbf{f}-\rho'\nabla \Phi \\
  & -\rho'\mathbf{A}-\rho_0\boldsymbol{\alpha}\times\mathbf{r}
    -2\rho_0\boldsymbol{\Omega}\times\mathbf{u}
    -\rho'\boldsymbol{\Omega}\times(\boldsymbol{\Omega}\times\mathbf{r}),
\end{aligned}
\end{equation}
where
\begin{equation}
  p^*=p+\rho_0\Phi-\frac{1}{2}\rho_0|\boldsymbol{\Omega}\times\mathbf{r}|^2
  +\rho_0\mathbf{A}\cdot\mathbf{r},
\end{equation}
together with the incompressibility condition
$\nabla\cdot\mathbf{u}=0$. Of course, supplementary equations are
often needed; for example, if $\rho'$ depends on the temperature, an
evolution equation for the temperature must be included.

\subsection{Formulation in the inertial frame}\label{sec:Bou_if}

In many cases the fluid container is not rotating at a given angular
speed, but different parts may rotate independently. For example
Taylor-Couette flows with stratification and/or heating, cylindrical
containers with the lids rotating at different angular velocities,
etc. In these flows, there is not a natural or unique angular velocity
$\boldsymbol{\Omega}$ to use in \eqref{Bou_rot_frame} and it may be
more convenient to write the governing equations in the laboratory
reference frame. In \S\ref{sec:Bou_if_Om} we derive the momentum
equation in the laboratory frame but for the sake of simplicity we
assume that the fluid container rotates with angular speed
$\mathbf{\Omega}$. In \S\ref{sec:Bou_if_Gen} we show how the
formulation is easily extended to account for the general case where a
unique rotating reference frame cannot be identified.

\subsubsection{Formulation in the inertial frame: container rotating
  at angular velocity $\mathbf{\Omega}$}\label{sec:Bou_if_Om}

The laboratory frame is an inertial reference frame, so the four
inertial terms in \eqref{NS1} are absent, and the momentum equation is
\begin{equation}\label{NS2}
  \rho(\partial_t+\mathbf{v}\cdot\nabla)\mathbf{v}= -\nabla p
    +\mu\nabla^2\mathbf{v}-\rho\nabla \Phi+\rho\,\mathbf{f},
\end{equation}
where we have used $\mathbf{v}$ for the velocity field in the inertial
reference frame, to distinguish it from the velocity $\mathbf{u}$ in
the rotating frame. In order to implement the Boussinesq
approximation, we could na\"ively repeat the previous analysis; since
the only term which is a gradient is the gravitational force
$-\rho_0\nabla \Phi$, we end up with an equation containing only the
gravitational buoyancy, and the centrifugal buoyancy is absent. This
appears reasonable, because the governing equations do not contain the
rotation frequency $\boldsymbol{\Omega}$ of the container. However,
$\boldsymbol{\Omega}$ appears in the boundary conditions for the
velocity, so it must be taken into account by a careful analysis of
the nonlinear advection term. The easiest way to do this is by
decomposing the velocity field as
$\mathbf{v}=\mathbf{u}+\boldsymbol{\Omega}\times\mathbf{r}$, so the
$\boldsymbol{\Omega}\times\mathbf{r}$ part accounts for the boundary
conditions (rotating container); $\mathbf{u}$ is precisely the
velocity of the fluid in the rotating reference frame, with zero
velocity boundary conditions. The advection term splits into four
parts:
\begin{equation}\label{advec1}
  \mathbf{v}\cdot\nabla\mathbf{v}= \mathbf{u}\cdot\nabla\mathbf{u}+
  \mathbf{u}\cdot\nabla(\boldsymbol{\Omega}\times\mathbf{r})+
  (\boldsymbol{\Omega}\times\mathbf{r})\cdot\nabla\mathbf{u}
  +(\boldsymbol{\Omega}\times\mathbf{r})\cdot\nabla
  (\boldsymbol{\Omega}\times\mathbf{r}).
\end{equation}
Using the incompressibility character of $\mathbf{u}$, the dependence
of $\boldsymbol{\Omega}$ on time but not on the spatial coordinates,
and some vector identities, we can transform the advection term into
\begin{equation}\label{advec2}
  \mathbf{v}\cdot\nabla\mathbf{v}=\mathbf{u}\cdot\nabla\mathbf{u}+
  2\,\boldsymbol{\Omega}\times\mathbf{u}+
  \boldsymbol{\Omega}\times(\boldsymbol{\Omega}\times\mathbf{r})
  +\nabla\times\big(\mathbf{u}\times(\boldsymbol{\Omega}\times\mathbf{r})\big).
\end{equation}
We have recovered the Coriolis and centrifugal terms, and because 
$\boldsymbol{\Omega}\times(\boldsymbol{\Omega}\times\mathbf{r})$ is a
gradient, we must add a centrifugal contribution also in the inertial
reference frame.

The last term in \eqref{advec2} accounts for the difference between
the time derivatives in the inertial and rotating reference frames
respectively. An easy way to see this is by considering the simple
case where the two reference frames have the same origin, and
$\boldsymbol{\Omega}=\Omega\hat{\mathbf{k}}$, where $\hat{\mathbf{k}}$
is the vertical unit vector and $\Omega$ is constant. Using
cylindrical coordinates $(r,\theta,z)$, with $z$ in the vertical
direction, we obtain
\begin{equation}\label{advec3}
  \nabla\times\big(\mathbf{u}\times(\boldsymbol{\Omega}\times\mathbf{r}))
  =\Omega\partial_\theta\mathbf{u}.
\end{equation}
The change of coordinates between the inertial and rotating frame is
\begin{equation}\label{coor_change}
\left.\begin{aligned}
 & r=r', & z=z', \\ & \theta=\theta'+\Omega t,\quad & t=t',
\end{aligned}~~\right\}
\end{equation}
where $(r',\theta',z')$ are the cylindrical coordinates in the
rotating frame of the same fluid parcel with coordinates
$(r,\theta,z)$ in the inertial frame; $t$ and $t'$ are the times in
both reference frames. From \eqref{coor_change} we obtain
$\partial_{t'}=\partial_t+\Omega\partial_\theta$, so the last term in
\eqref{advec2}, combined with $\partial_t\mathbf{u}$ results in the
term $\partial_{t'}\mathbf{u}$ in the rotating frame. Finally,
$\partial_t\mathbf{v}$ in the inertial frame contains an extra term,
$\partial_t(\boldsymbol{\Omega}\times\mathbf{r})=
\boldsymbol{\alpha}\times\mathbf{r}$. Therefore, we have recovered all
the inertial forces in the rotating frame \eqref{NS1}, except for the
translation force $-\rho\mathbf{A}$, because in the example
considered, \eqref{coor_change}, both reference frames have the same
origin, and the translation is absent; by including a translation term
in \eqref{coor_change} we could also recover it. Now, the two
formulations, including centrifugal buoyancy in both reference frames
(inertial and rotating), fully agree.

In the inertial reference frame, we are interested in a formulation in
terms of the velocity field in the inertial frame $\mathbf{v}$,
instead of $\mathbf{u}$ as in \eqref{advec2}. The analysis presented
above considering the advection term results simply in an additional
term, the centrifugal buoyancy. We have also discussed the effect of
the decomposition
$\mathbf{v}=\mathbf{u}+\boldsymbol{\Omega}\times\mathbf{r}$ in the
time derivative term. Now it only remains to consider the viscous
term. However, $\nabla^2(\boldsymbol{\Omega}\times\mathbf{r})=0$
because $\boldsymbol{\Omega}\times\mathbf{r}$ is linear in the spatial
coordinates and so its Laplacian is zero. The traditional Boussinesq
approximation equations in the inertial reference frame are
\begin{equation}\label{Bou_iner_frame}
  \rho_0(\partial_t+\mathbf{v}\cdot\nabla)\mathbf{v}= -\nabla p^*
  +\mu\nabla^2\mathbf{v}+\rho\,\mathbf{f}-\rho'\nabla \Phi
  -\rho'\boldsymbol{\Omega}\times(\boldsymbol{\Omega}\times\mathbf{r}),
\end{equation}
where $p^*=p+\rho_0\Phi-\frac{1}{2}\rho_0|\boldsymbol{\Omega}\times
\mathbf{r}|^2$, and together with the incompressibility condition
$\nabla\cdot\mathbf{u}=0$.

\subsubsection{Formulation in the inertial frame:
  generalization}\label{sec:Bou_if_Gen}

We have shown that centrifugal buoyancy enters the governing equations
via the boundary conditions and the advection term; no other term is
affected in the Boussinesq approximation. This now suggests a very
simple formulation, consisting in keeping the whole density,
$\rho=\rho_0+\rho'$, in the advection term. This formulation is easy
to implement, and since most time-evolution codes for incompressible
flows are semi-implicit (i.e.\ the viscous term is treated implicitly,
whereas the advection term is treated explicitly), the speed and
efficiency of the codes do not change. The formulation reads
\begin{equation}\label{Bou_iner_frame_gen}
  \rho_0(\partial_t+\mathbf{v}\cdot\nabla)\mathbf{v}=
    -\nabla p^* +\mu\nabla^2\mathbf{v}+\rho\,\mathbf{f}-\rho'\nabla\Phi
    -\rho'(\mathbf{v}\cdot\nabla)\mathbf{v},
\end{equation}
where $p^*=p+\rho_0\Phi$, and allows one to easily handle situations
where different parts of a fluid container rotate independently. In
these flows there is not a natural or unique angular velocity
$\boldsymbol{\Omega}$ to use for a rotating reference frame in the
formulation \eqref{Bou_iner_frame}; however the angular velocities of
the problem still enter the governing equations through the boundary
conditions and the advection term. Hence formulation
\eqref{Bou_iner_frame_gen} provides a natural way to account for
centrifugal buoyancy effects of these rotating flows in the inertial
(laboratory) reference frame. This formulation is also appropriate if
additional equations appear coupled with the Navier-Stokes equations,
for example for large density variations in stratified flows. The
treatment of the centrifugal effects can be carried out exactly in the
same way presented here.

\subsubsection{Alternative formulation in the inertial frame and
  physical interpretation}\label{sec:Bou_if_Alt}

The extra term included in \eqref{Bou_iner_frame_gen},
$\rho'(\mathbf{v}\cdot\nabla)\mathbf{v}$, can be expressed in a
different way, providing a closer resemblance to the expression in
\eqref{Bou_iner_frame}. Close to a rotating wall, the velocity field
is $\mathbf{v}\approx\boldsymbol{\Omega}\times\mathbf{r}$; this
expression is exact at the wall (no slip boundary condition at a rigid
rotating wall). The dominant part of the advection term is then
\begin{equation}
  (\mathbf{v}\cdot\nabla)\mathbf{v}\approx(\boldsymbol{\Omega}\times
    \mathbf{r})\cdot\nabla(\boldsymbol{\Omega}\times\mathbf{r})=
    \boldsymbol{\Omega}\times(\boldsymbol{\Omega}\times\mathbf{r}) 
  =-\nabla(\frac{1}{2}|\boldsymbol{\Omega}\times\mathbf{r}|^2)
    \approx-\nabla\big(\,\frac{1}{2}\mathbf{v}^2\big).
\end{equation}
As the dominant term is a gradient, it is necessary to include the
$\rho'$ term in the Boussinesq approximation.
Replacing $\rho'(\mathbf{v}\cdot\nabla)\mathbf{v}$ by
$-\rho'\nabla(\frac{1}{2}\mathbf{v}^2)$ gives the
alternative form for \eqref{Bou_iner_frame_gen}:
\begin{equation}\label{Bou_iner_frame_gen_bo}
  \rho_0(\partial_t+\mathbf{v}\cdot\nabla)\mathbf{v} = 
  -\nabla p^* +\mu\nabla^2\mathbf{v}+\rho\,\mathbf{f}
  -\rho'\nabla \Phi +\rho'\nabla\big(\,\frac{1}{2}\mathbf{v}^2\big).
\end{equation}
This centrifugal effect is not only important when we have rotating
walls, but also if a strong vortex appears dynamically in the interior
of the domain; therefore it is advisable to always include this term
in the Boussinesq approximation in order to account for all possible
sources of centrifugal instability.

We have presented two different ways, \eqref{Bou_iner_frame_gen} and
\eqref{Bou_iner_frame_gen_bo}, of including the centrifugal buoyancy
in rotating problems. One may wonder if there exists a canonical way
to extract from the advection term the part that is a gradient, and
then multiply this gradient by $\rho'$. The Helmholtz decomposition
\citep{ArWe01}, writing a given vector field as the sum of a gradient
and a curl, could serve this purpose, but unfortunately this
decomposition is not unique (it depends on the boundary conditions
satisfied by the curl part), and moreover it is not a local
decomposition (i.e., in order to extract the gradient part, we need to
solve a Laplace equation with Neumann boundary conditions). The two
formulations presented here, \eqref{Bou_iner_frame_gen} and
\eqref{Bou_iner_frame_gen_bo}, are simple and easy to implement, and
deciding between one or the other is a matter of taste.

The extra term we have included in \eqref{Bou_iner_frame_gen_bo}, 
$\rho'\nabla(\frac{1}{2}\mathbf{v}^2)$, has an important physical
interpretation; it is a source of vorticity due to density variations and
centrifugal effects. Taking the curl of \eqref{Bou_iner_frame_gen_bo}
and using
\begin{equation}
  \nabla\times(\mathbf{v}\cdot\nabla\mathbf{v})=
  \nabla\times(\boldsymbol{\omega}\times\mathbf{v})=
  \mathbf{v}\cdot\nabla\boldsymbol{\omega}-
  \boldsymbol{\omega}\cdot\nabla\mathbf{v},
\end{equation}
where $\boldsymbol{\omega}=\nabla\times\mathbf{v}$ is the vorticity
field, results in an equation for the vorticity:
\begin{equation}\label{Bou_iner_frame_vort}
  \rho_0(\partial_t+\mathbf{v}\cdot\nabla)\boldsymbol{\omega}=
  \rho_0\boldsymbol{\omega}\cdot\nabla\mathbf{v}+\mu\nabla^2\boldsymbol{\omega}
  +\nabla\times(\rho\,\mathbf{f})-\nabla\rho'\times\nabla \Phi
  +\nabla\rho'\times\nabla\big(\,\frac{1}{2}\mathbf{v}^2\big).
\end{equation}
The first three terms in the right-hand-side of
\eqref{Bou_iner_frame_vort} provide the classical vorticity evolution
equation for an incompressible flow with constant density. The last
two terms are the explicit generation of vorticity due to the
gravitational and centrifugal buoyancies, respectively.  In the next
section we discuss two hydrodynamic approaches to the accretion disk
problem in astrophysics, where centrifugal buoyancy is not included,
and we show that it can be easily included in the numerical analysis.

\section{Centrifugal effects in hydrodynamic accretion disk models}

There are other approximations used in the literature, which may be
also modified to include centrifugal buoyancy. Astrophysics is a very
active field where these approximations are used. The book of
\citet{Tass00} provides a comprehensive discussion on rotating stellar
flows under the influence of shear and stratification. In this section
we briefly discuss two approximations used in accretion disk
theory. The first is the shearing sheet model
\citep{Bal03,ReUm08,LePa10}, where the Boussinesq approximation is
used in a small domain of the accretion disk. The second is the
anelastic approximation \citep{Ban96}, used by \citet{PJS07} in a
global model of an accretion disk.

In the shearing sheet approximation the governing equations are
written in a small thin rectangular box at a distance $r_0$ from the
centre of the accretion disk; the coordinates used are $x=r-r_0$,
$y=r_0\theta$, and $z$, where $(r,\theta,z)$ are the cylindrical polar
coordinates of the disk. Let $\Omega(r)$ be the Keplerian angular
velocity profile of the accretion disk, i.e.\ its background
rotation. The rotating reference frame has $\boldsymbol{\Omega}=
\Omega_0{\bf e}_z$, $\mathbf{A}=-r_0\Omega_0^2{\bf e}_r$ and
$\boldsymbol \alpha=0$, where $\Omega_0=\Omega(r_0)$ (see
\ref{Bou_rot_frame}). In terms of the velocity perturbation with
respect to the background rotation,
$\mathbf{w}=(u,v,w)=\mathbf{u}-\mathbf{u}_0$, with $\mathbf{u}_0=
r(\Omega(r)-\Omega_0){\bf e}_y$, the governing equations
\eqref{Bou_rot_frame} are
\begin{equation}\label{SSB}
  \begin{aligned}
    \rho_0(\partial_t+ & \mathbf{w}\cdot\nabla-Sx\partial_y)\mathbf{w}=
      -\nabla p^*+\mu\nabla^2\mathbf{w}-\rho'\nabla \Phi \\
    & -2\rho_0\boldsymbol{\Omega}\times\mathbf{w}
      +\rho_0Su{\bf e}_y-2\rho_0\Omega_0Sx{\bf e}_x-\rho'\boldsymbol{\Omega}
      \times\big(\boldsymbol{\Omega}\times(r_0{\bf e}_x+\mathbf{r})\big).
  \end{aligned}
\end{equation}
Here $S=-r_0d\Omega/dr|_{r_0}$ is a linear approximation of the shear
associated with the background rotation profile $\Omega(r)$. We have
assumed as customary that $x\ll r_0$ and expanded $\Omega(r)$ up to
first order in $x/r_0$. We can compare \eqref{SSB} with the governing
equations in \citet{Bal03,LePa10}, and we observe that the centrifugal
therm $-\rho'\boldsymbol{\Omega}\times(\boldsymbol{\Omega}\times
\mathbf{r})$ is absent in these references.  The baroclinic term
$-\rho'\nabla \Phi$ is the only buoyancy term considered in these
works, and it points into the radial direction for an axisymmetric
mass distribution in the accretion disk. Another source of instability
are the shear terms proportional to $S$, that are independent of the
temperature. When centrifugal buoyancy is included, additional terms
both in the radial and azimuthal directions appear, competing with the
gravitational buoyancy and the shear terms. As a result, the stability
analysis and the dynamics of the accretion disk may be modified by the
inclusion of centrifugal buoyancy. If the centrifugal effects of
internal strong vortices or differential rotation are also taken into
account, like in (\ref{Bou_iner_frame_gen},
\ref{Bou_iner_frame_gen_bo}), additional terms may also be included:
$-\rho'(\mathbf{v}\cdot\nabla)\mathbf{v}$ or equivalently
$\rho'\nabla\big(\,\frac{1}{2}\mathbf{v}^2\big)$.

The shearing sheet approximation is local, it models a small
rectangular neighbourhood of a point in the accretion disk. In order
to perform a global analysis of the disk in the radial direction, it
is necessary to account for large variations in density, which do not
fit into the Boussinesq framework. The anelastic approximation is very
useful in this case. It is assumed that there is a background state
$\rho_0(r)$, $p_0(r)$ in static balance between centrifugal force,
gravity and pressure,
\begin{equation}\label{statbal}
  r\Omega^2(r)=\frac{d\Phi}{dr}+\frac{1}{\rho_0}\frac{dp_0}{dr},
\end{equation}
and the continuity equation now reads
$\nabla\cdot(\rho_0(r)\mathbf{u})=0$. The velocity field is not
solenoidal, but the governing equations and numerical methods are very
similar to those corresponding to the Navier-Stokes-Boussinesq
approximation, and in 2D problems \citep{PJS07} a streamfunction can
still be defined. Because of the strong differential rotation in the
accretion disk problem, the inertial reference frame is usually
preferred. As the centrifugal force is included in the static balance
\eqref{statbal}, it may look like centrifugal effects have been
included into the governing equations. However, the static balance
means that the centrifugal term
$-\rho_0\boldsymbol{\Omega}\times(\boldsymbol{\Omega}\times
\mathbf{r})$ is a gradient, and therefore terms of the form $-\rho'
(\mathbf{v}\cdot\nabla)\mathbf{v}$ or
$\rho'\nabla\big(\,\frac{1}{2}\mathbf{v}^2\big)$ should be included in
the governing equations, as has been discussed in the preceding
section. These terms are not included in studies using the anelastic
approximation \citep{Ban96,PJS07}. Therefore centrifugal effects in
many geophysical and astrophysical problems could modify the stability
analysis and the dynamics obtained so far, particularly at large
rotation rates.

%%%%%%%%%%%%%%%%%%%%%%%%%%%%%%%%%%%%%%%%%%%%%%%%%%%%%%%%%%%%%%%%%%%%%%

\section{Description of the system}

We consider the motion of a fluid of kinematic viscosity $\nu$
contained in the annular gap between two concentric infinite cylinders
of radii $r_i$ and $r_o$. The cylinders rotate at independent angular
speeds $\Omega_i$ and $\Omega_o$. A negative radial gradient of
temperature, as in accretion disks, is considered by setting the
temperature of the inner cylinder to $T_i=T_c+\Delta T/2$ and the
outer cylinder to $T_o=T_c-\Delta T/2$, where $T_c$ is the mean
temperature. We fix the radii ratio $\eta=0.71$, a typical value in
experimental facilities, and the Prandtl number $\sigma=7.16$,
corresponding to water. In astrophysics $\sigma\ll 1$ because thermal
relaxation is dominated by radiation processes, whereas in geophysics
(planetary core and mantle) $\sigma\gg 1$. We assume that the
gravitational acceleration is vertical and uniform, as in typical
Taylor-Couette experiments. This is in contrast to astrophysical
stellar flows, where radial gravity plays a prominent role and cannot
be neglected \citep{Tass00}. For example, the radial buoyancy
frequency (absent in our system) defines the stability of rotating
astrophysical objects. Similarly, in accretion disks the radial
Grashof number (also absent in our system) is more relevant than the
vertical one. Another crucial difference is the presence of radial
boundaries (cylinders) to drive rotation. As a result, in the
quasi-Keplerian regime ($\Omega_i>\Omega_o$ and
$r_i^2\Omega_i<r_o^2\Omega_o$) the radial pressure gradient is
positive, whereas in accretion disks it may also be negative.

\subsection{Governing equations}

The centrifugal buoyancy in the stationary frame of
reference is included as in \S2, equation\eqref{Bou_iner_frame_gen}:
\begin{equation}\label{formII_1}
  \rho_0(\partial_t+\mathbf{v}\cdot\nabla)\mathbf{v}=-\nabla p^*
  +\mu\nabla^2\mathbf{v}-\rho'\nabla\Phi
  -\rho'\mathbf{v}\cdot\nabla\mathbf{v},
\end{equation}
where $p^*$ includes part of the gravitational potential, $\rho_0\Phi$.

We assume $\rho=\rho_0+\rho'=\rho_0(1-\alpha T)$, where $T$ is the
deviation of the temperature with respect to the mean temperature
$T_c$, and $\rho_0$ is the density of the fluid at $T_c$. As the
gravity acceleration is vertical and uniform, the gravitational
potential is given by $\Phi=gz$; cylindrical coordinates
$(r,\theta,z)$ are used. With these assumptions,
$-\rho'\nabla\Phi=\rho_0\alpha gT{\bf\hat z}$ where ${\bf\hat z}$ is
the unit vector in the axial direction $z$ and $\alpha$ is the
coefficient of volume expansion. The governing equations, including
the temperature and incompressibility condition, are:
\begin{subequations}\label{formII}
  \begin{align}
    & (\partial_t+\mathbf{v}\cdot\nabla)\mathbf{v}=-\nabla p
      +\nu\nabla^2\mathbf{v}+\alpha gT{\bf\hat z}
      +\alpha T\mathbf{v}\cdot\nabla\mathbf{v},\label{formII_2v} \\
    & (\partial_t+\mathbf{v}\cdot\nabla)T=\kappa\nabla^2T,\label{formII_2T}\\
    & \nabla\cdot\mathbf{v}=0,\label{formII_2inc}
  \end{align}
\end{subequations}
where $\kappa$ is the thermal diffusivity of the fluid.  The equations
are made dimensionless using the gap width $d=r_o-r_i$ as the length
scale, the viscous time $d^2/\nu$ as the time scale, $\Delta T$ as the
temperature scale, and $(\nu/d)^2$ for the pressure. In doing so, six
independent dimensionless numbers appear:
\begin{subequations}\label{nondim_numbers}
  \begin{alignat}{3}
    & \text{Grashof number} && G=\alpha g\Delta Td^3/\nu^2, \\
    & \text{relative density variation} &\qquad&
      \epsilon=\alpha\Delta T=\Delta\rho/\rho_0, \\
    & \text{Prandtl number} && \sigma=\nu/\kappa, \\
    & \text{radius ratio} && \eta=r_i/r_o, \\
    & \text{inner Reynolds number} && \Rey_i=\Omega_ir_id/\nu, \\
    & \text{outer Reynolds number} && \Rey_o=\Omega_or_od/\nu.
  \end{alignat}
\end{subequations}
where $\Delta\rho$ is the density variation associated with a
temperature change of $\Delta T$. In this system the Froude number is
not particularly useful because we have two different rotation rates,
$\Omega_i$ and $\Omega_o$, so the Froude number definition is not
unique.

From now on, only dimensionless variables and parameters will be used.
The dimensionless governing equations are:
\begin{subequations}\label{formIInon}
\begin{align}
  & (\partial_t+\mathbf{v}\cdot\nabla)\mathbf{v}=-\nabla p
    +\nabla^2\mathbf{v}+G T{\bf\hat z}
    +\epsilon T\mathbf{v}\cdot\nabla\mathbf{v},\label{formII_3v}\\
  & (\partial_t+\mathbf{v}\cdot\nabla)T=\sigma^{-1}\nabla^2T,\label{formII_3T}\\
  & \nabla\cdot\mathbf{v}=0,\label{formII_3inc}
\end{align}
\end{subequations}
The only change needed to recover the traditional Boussinesq
approximation is to replace the centrifugal term $\epsilon
T\mathbf{v}\cdot\nabla\mathbf{v}$ in \eqref{formII_3v} by $-\epsilon
\Omega^2 T r {\bf\hat r}$, where ${\bf\hat r}$ is the unit vector in
the radial direction $r$.

\subsection{Base flow}

An analytical solution for the base flow can be found by assuming
only radial dependence for the variables of the problem. We also use
the zero axial mass flux condition to fix the axial pressure gradient, i.e.:
\begin{equation}\label{zero_axial_flux}
  \int^{r_o}_{r_i} rw_b(r) dr = 0.
\end{equation}
The resulting steady base flow is given by:
\begin{subequations}\label{basicflow}
  \begin{align}
    & u_b(r)=0 \label{ur} \\
    & v_b(r)=Ar+\frac{B}{r} \label{ut} \\
    & w_b(r)=G\biggr(C(r^2-r_i^2)+\Big(C(r_o^2-r_i^2)+\frac{1}{4}(r_o^2-r^2)
      \Big)\frac{\ln(r/r_i)}{\ln\eta}\biggr) \label{uz} \\
    & T_b(r)= \frac{1}{2}+\frac{\ln(r/r_i)}{\ln\eta} \label{tem} \\
    & p(r,z)=p_o+G\Big(4C+\frac{1}{2}-\frac{1}{\ln\eta}\Big)z+\int_{r_i}^r
      \big(1-\epsilon T_b(r)\big)v_b^2(r)\frac{dr}{r}. \label{basic_pressure}
  \end{align}
\end{subequations}
where $(u,v,w)$ are the radial, azimuthal and axial components of the
velocity field, and cylindrical coordinates $(r,\theta,z)$ are being used.
$v_b$ is the azimuthal velocity for the classical Taylor-Couette
problem \citep{Ch61}, whereas $w_b$ and $T_b$ correspond to convection
in a conductive regime and appeared for the first time in
\citet{ChKp80}. The pressure varies linearly with the axial coordinate
$z$, but the pressure gradient depends only on $r$, and therefore it
is periodic in the axial direction. This axial pressure gradient
mimics the presence of distant endwalls in any real situation, by
enforcing the zero mass flux constraint \eqref{zero_axial_flux}. It is
possible to give an explicit closed expression for $p$ by integrating
\eqref{basic_pressure}, but it is quite involved and it does not
appear in the problem solution. The expressions for the parameters
$A$, $B$ and $C$ are:
\begin{align}
  & A=\frac{\Rey_o-\eta \Rey_i}{1+\eta}, \qquad
    B=\eta \frac{\Rey_i-\eta \Rey_o}{(1-\eta)(1-\eta^2)}, \label{param_rot} \\
  & C=-\frac{4\ln\eta+(1-\eta^2)(3-\eta^2)}
    {16(1-\eta^2)\big((1+\eta^2)\ln\eta+1-\eta^2\big)},
\end{align}
where \eqref{param_rot} define the pure rotational flow in the
azimuthal coordinate and $C$ gives the axial component of the velocity
field. The non-dimensional radii of the cylindrical walls are given by
$r_i=\eta/(1-\eta)$, $r_o=1/(1-\eta)$. Note that the presence of the
new centrifugal buoyancy term, proportional to $\epsilon$, does not
modify the base flow's velocity field, but only its pressure.

\subsection{Linearized equations}

We perturb the base flow with infinitesimal perturbations which vary
periodically in the axial and azimuthal directions,
\begin{subequations}\label{perturbed}
  \begin{align}
    & \mathbf{v}(r,\theta,z,t)=\mathbf{v_b}(r)+e^{i(n\theta+kz)+\lambda t}
      \mathbf{u}(r), \label{vper} \\
    & T(r,\theta,z,t)=T_b(r)+e^{i(n\theta+kz)+\lambda t}
      T^\prime(r), \label{Tper}
   \end{align}
\end{subequations}
where $\mathbf{v_b}=(0,v_b,w_b)$ and $T_b(r)$ correspond to the base
flow \eqref{basicflow}; $\mathbf{u}(r)=(u_r,u_\theta,u_z)$ and
$T^\prime(r)$ are the velocity and temperature perturbations,
respectively. The boundary conditions for both $\mathbf{u}$ and
$T^\prime$ are homogeneous: $\mathbf{u}(r_i)=
\mathbf{u}(r_o)=T^\prime(r_i)=T^\prime(r_o)=0$. The axial wavenumber
$k$ and the azimuthal mode number $n$ define the shape of the
disturbance. The parameter $\lambda$ is complex. Its real part
$\lambda_r$ is the perturbation's growth rate, which is zero at
critical values, and its imaginary part $\lambda_i$ is the oscillation
frequency of the perturbation.

Using the decomposition \eqref{perturbed} in the equations
\eqref{formIInon} and neglecting high-order terms, we obtain an
eigenvalue problem, with eigenvalue $\lambda$. It reads
\begin{subequations}\label{lineq}  
  \begin{alignat}{3}
    & \lambda u_r &\;=\;& \frac{1}{r}\frac{\partial}{\partial r}
      (r\frac{\partial u_r}{\partial r})-u_r[\frac{n^2+1}{r^2}+k^2
      +i(\frac{nv_b}{r}+kw_b)(1-\epsilon T_b)] \nonumber \\
    & && +\frac{2v_b}{r}(1-\epsilon T_b)u_\theta-\frac{2in}{r^2}u_\theta
         -\frac{\epsilon v_b^2}{r}T', \\
    & \lambda u_\theta &\;=\;& \frac{1}{r}\frac{\partial}{\partial r}
      (r\frac{\partial u_\theta}{\partial r})-u_\theta[\frac{n^2+1}{r^2}
      +k^2+i(\frac{nv_b}{r}+kw_b)(1-\epsilon T_b)] \nonumber \\
    & && -(\frac{\partial v_b}{\partial r}+\frac{v_b}{r})(1-\epsilon T_b)
         u_r+\frac{2in}{r^2}u_r, \\
    & \lambda u_z &\;=\;& \frac{1}{r}\frac{\partial}{\partial r}
      (r\frac{\partial u_z}{\partial r})-u_z[\frac{n^2}{r^2}+k^2
      +i(\frac{nv_b}{r}+kw_b)(1-\epsilon T_b)] \nonumber \\
    & && +\frac{\partial w_b}{\partial r}(\epsilon T_b-1)u_r+GT', \\
    & \lambda T' &\;=\;& \frac{1}{\sigma r}\frac{\partial}{\partial r}
      (r\frac{\partial T'}{\partial r})-T'[\frac{1}{\sigma}
      (\frac{n^2}{r^2}+k^2)+i(\frac{nv_b}{r}+kw_b)]
      -\frac{\partial T_b}{\partial r}u_r.
  \end{alignat}
\end{subequations}
Note that here the continuity equations and pressure terms are omitted
because the Petrov--Galerkin method chosen to solve the resulting
system of equations automatically satisfies the continuity equation
and eliminates the pressure by using a proper projection (see next
section).

From \eqref{lineq} the equations for the traditional Boussinesq
approximation can be easily obtained by setting
$\epsilon=0$ in all terms except for $-\epsilon(v_b^2/r)T'$. The
traditional approximation incorporates only one rotating frame of
reference for the system; the expression \eqref{ut} for the base flow
azimuthal velocity $v_b(r)=Ar+B/r$ has two terms, $Ar$ corresponding
to solid body rotation, and $B/r$ corresponding to shear. It is
natural to identify $A$ as the frequency of the rotating frame of
reference, $\Omega_r$. In fact, if we take $\Omega_i=\Omega_o=\Omega$,
the Couette flow profile is:
\begin{equation}\label{TCOmega}
  v_b(r)=Ar+\frac{B}{r}=\frac{\Omega_o r_o^2-\Omega_ir_i^2}{r_o^2-r_i^2}r+
  \frac{(\Omega_i-\Omega_o)(r_ir_o)^2}{r_o^2-r_i^2}\frac{1}{r}=\Omega r
  =\Omega_rr,
\end{equation}
and we recover the linearized version of the centrifugal term
consisered in the traditional approach, $-\epsilon \Omega^2T^\prime
r{\bf\hat r}$. In the general case with $\Omega_i\ne\Omega_o$ the
traditional Boussinesq approximation is defined in the frame of
reference rotating with $\Omega_r=A$. This approximation takes only
into account the centrifugal buoyancy acting in the radial direction,
which is obviously its main contribution. However, as we will see in
\S\ref{Sec:results}, for high rotation rates other terms acting both
in the radial and azimuthal directions become important and change the
behaviour of the system. Part of the discrepancy stems from the fact
that the effect of differential rotation is entirely neglected in the
traditional approximation.

%%%%%%%%%%%%%%%%%%%%%%%%%%%%%%%%%%%%%%%%%%%%%%%%%%%%%%%%%%%%%%%%%%%%%%    

\section{Numerical method}

In order to solve numerically the eigenvalue problem described in the
previous section, a spatial discretization of the domain must be made.
This is accomplished by projecting the equations \eqref{lineq} onto 
a basis carefully chosen to simplify the process,
\begin{equation}\label{vectsp}
  V_3=\{\mathbf{v}\in (\mathscr{L}_2(r_i,r_o))^3\ |\ \nabla\cdot
  \mathbf{v}=0,\ \mathbf{v}(r_i)=\mathbf{v}(r_o)=0\},
\end{equation}
where $(\mathscr{L}_2(r_i,r_o))^3$ is the Hilbert space of
square integrable vectorial functions defined on the interval
($r_i,r_o$), with the inner product
\begin{equation}
  \langle\mathbf{v},\mathbf{u}\rangle=\int_{r_i}^{r_o}\mathbf{v}^*
  \cdot\mathbf{u}\ rdr,
\end{equation}
where * denotes the complex conjugate. For any $\mathbf{v} \in V_3$
and any function $p$, using the incompressibility condition, the
boundary conditions and integrating by parts,
\begin{equation}
  \langle\mathbf{v},\nabla p\rangle=\int_{r_i}^{r_o}(\mathbf{v}^*\cdot\nabla p)
  rdr=\int_{r_i}^{r_o}rv^*_r\partial_rp\,dr=rpv^*_r|_{r_i}^{r_o}-\int_{r_i}^{r_o}p
  \partial_r(rv^*_r)dr=0.
\end{equation}
This consideration allows us to eliminate the pressure from the
equations as we project them onto the basis \citep{CQHZ07}. Moreover, the
continuity equation is satisfied by definition of the space $V_3$. For
the temperature perturbation the appropriate space is
\begin{equation}\label{tempsp}
  V_1=\{f\in \mathscr{L}_2(r_i,r_o)\ |\ f(r_i)=f(r_o)=0\}.
\end{equation}
We expand the variables of the problem as follows
\begin{equation}\label{expansion}
  \mathbf{X}=\left[\begin{array}{c}\mathbf{u}(r)\\T'(r)\end{array}
  \right]=\sum_j a_j\mathbf{X}_j\qquad\mathbf{X}_j\in
  V_3\times V_1,
\end{equation}
and projecting \eqref{lineq} onto $V_3\times V_1$, we arrive at a linear
system of equations for the coefficients $a_j$.

The solution of the system is performed by means of a Petrov-Galerkin
scheme, where the basis used in the expansion is different from the
one used in the projection. The bases are composed of functions built
on Chebyshev polynomials satisfying the boundary conditions. A
detailed description of the method as well as the basis and functions
used for the velocity field can be found in \citet{MeMa00} and
\citet{MAMM07}, respectively. The basis functions for the
temperature (last component of ${\bf X}_j$ in \ref{expansion}),
and for the projection (with $\tilde{~}\,$) are:
\begin{equation}
  h_j(r)=(1-y^2)T_{j-1}(y),\qquad \tilde h_j(r)=r^2(1-y^2)T_{j-1}(y),
\end{equation}
where $y=2(r-r_i)-1$ and $T_j$ are the Chebyshev polynomials. As
a result of this process, we obtain a generalized eigenvalue system of
the form
\begin{equation}
  \lambda M_1 x = M_2 x,
\end{equation}
where $x$ is a vector containing the complex spectral
coefficients($a_j$) and the matrices $M_1$ and $M_2$ depend on the
parameters of the problem, the axial wavenumber $k$ and the azimuthal
mode $n$. This system is solved by using \emph{LAPACK}. The numerical
code written to perform this work implements the described method and
analyses a range of $k$, $n$ and $G$ provided by the user for a fixed
$\Rey$ number, searching for the critical values
($\Re\lambda=\lambda_r=0$). Up to $M=200$ radial modes have been used
in order to ensure the spectral convergence when high $\Rey$ numbers
are considered. The code has been tested by computing critical values
for several cases in \citet{McCoBo84} and \citet{AlWe90}, obtaining an
excellent agreement with their results, as shown in
table~\ref{codetesting}: the critical values computed coincide up to
the last digit shown with those in the mentioned references. In both
cases the outer cylinder is at rest ($\Rey_o=0$).

\begin{table}
  \begin{center}
    \begin{tabular}{cccc|ccccc}
      & \multicolumn{3}{c|}{Parameters} & \multicolumn{5}{c}{Critical
        values} \\\hline 
      & $\eta$ & $\sigma$ & $Ta$     & $G_c$    & $k_c$  & $n_c$ &
        $\lambda_i$ & $c=|\lambda_i|/k_c$ \\\hline 
      $(a)$  & $0.99$ & $0.71$   & 0        & $8038.0$ & $2.80$ & $~0$
        &  & $0.25424$   \\ 
      & $0.60$ & $0.71$   & 0        & $8512.4$ & $2.75$ & $~0$   &  &
        $13.39899$  \\ 
      & $0.60$ & $3.5$    & 0        & $8347.5$ & $2.75$ & $~0$   &  &
        $12.97744$  \\ 
      & $0.99$ & $3.5$    & 0        & $7857.1$ & $2.75$ & $~0$   &  &
        $0.24673$   \\\hline 
      $(b)$  & $0.6$  & $4.35$   & $2591.0$ & $50.0$   & $3.15$ & $~0$
        & $-0.50294$  &  \\ 
      & $0.6$  & $4.35$   & $380.3$  & $700.0$  & $1.88$ & $-3$  &
        $18.66889$  &  \\ 
      & $0.6$  & $15$     & $111.1$  & $280.0$  & $1.68$ & $-2$  &
        $6.58916$   &  \\  
      & $0.6$  & $15$     & $26.88$  & $700.0$  & $0.77$ & $-4$  &
        $7.19973$   &  \\\hline 
    \end{tabular}
    \caption{Code testing. The cases computed correspond to parameter
      values in $(a)$ \citet[table 1]{McCoBo84}, and $(b)$
      \citet[table 1, pg 67]{AlWe90}. $Ta=2(1-\eta)\Rey_i/(1+\eta)$ is
      the Taylor number, $\lambda_i=\Im[\lambda]$ is the imaginary
      part of the critical eigenvalue, for which $\Re[\lambda]=0$, and
      $c$ is the dimensionless axial wave speed. The sign of $n_c$ in
      our computation is opposite to that of \citet{AlWe90} because
      of the defintion of the normal Fourier modes in \eqref{perturbed}.}
  \end{center}
  \label{codetesting}
\end{table}

%%%%%%%%%%%%%%%%%%%%%%%%%%%%%%%%%%%%%%%%%%%%%%%%%%%%%%%%%%%%%%%%%%%%%%

\section{Stability of differentially heated fluid between co-rotating
  cylinders}\label{Sec:results}

In this section we present a detailed comparison of the linear
stability of the system using the traditional Boussinesq approximation
and the new approximation \eqref{lineq}. We consider three different
cases, all with $\eta = 0.71$ and $\sigma=7.16$. In the first one the
cylinders are rotating at same angular speed, corresponding to fluid
rotating as a solid-body. In the second and third cases the stability
of a differentially rotating fluid is considered in the presence of
weak and strong (quasi-Keplerian) shear.

\subsection{Cylinders rotating at same angular speed}\label{solid}

\begin{figure}
  \begin{center}
    \includegraphics[width=0.6\linewidth]{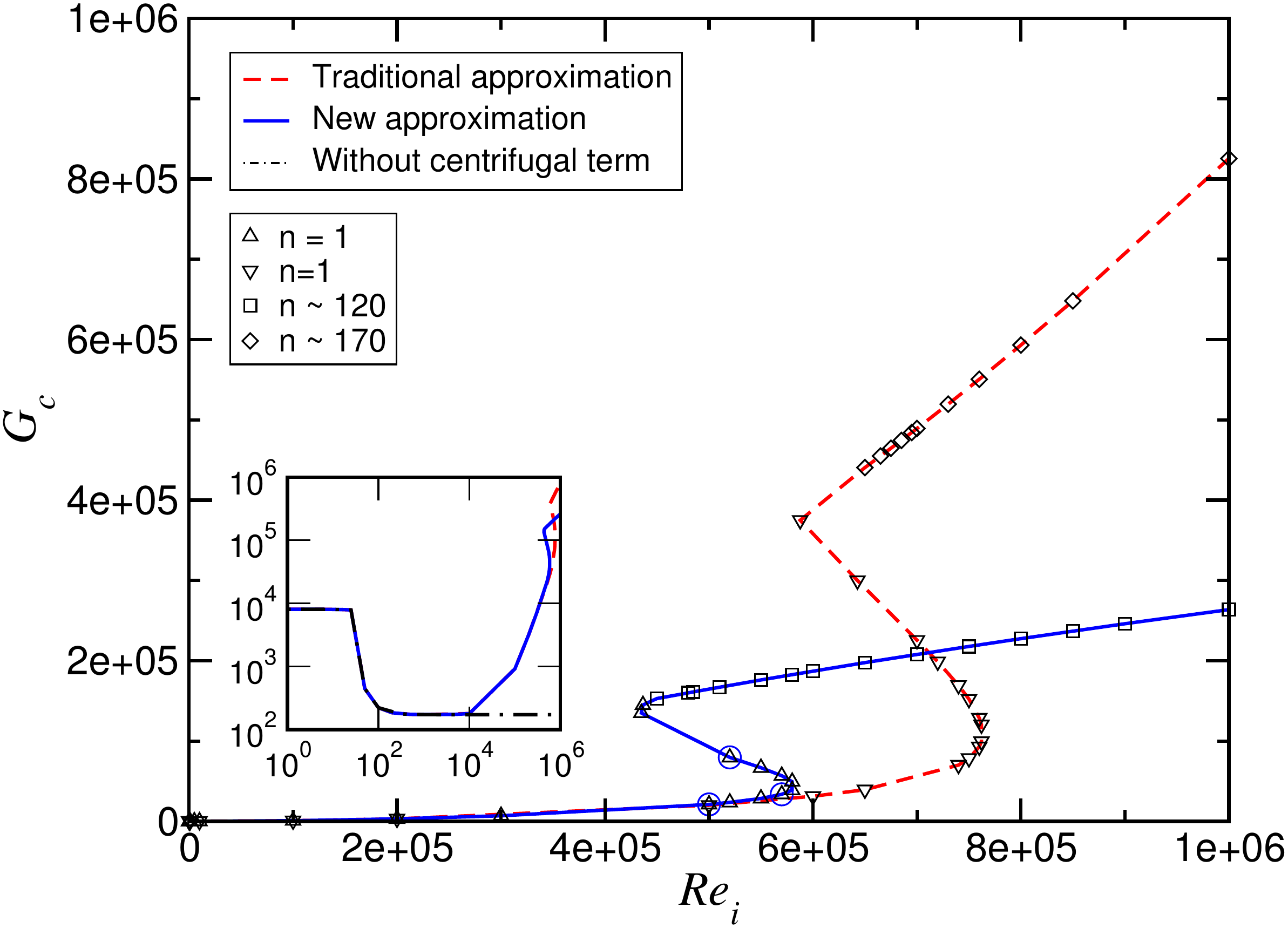}
  \end{center}
  \caption{Critical Grashof number $G_c$ as function of inner cylinder
    Reynolds number $\Rey_i$ for fluid rotating as a solid-body. The
    solid line is the linear stability curve using the new
    approximation for the centrifugal bouyancy proposed in this paper,
    the dashed corresponds to the traditional Boussinesq
    approach, whereas the dotted-dashed line is the case without centrifugal
    buoyancy, which can only be distinguished from the horizontal axis in the
    inset (log-log axes).  
    Different symbols indicate the two distinct mechanisms
    of instability. Up and down triangles represent the critical
    points due to the mechanism at moderate $Re_i$ for the new and
    traditional approximations respectively, whereas squares and
    diamonds correspond to the mechanism at large $Re_i$.}
  \label{Gr_Re_nosh}
\end{figure}  

\begin{figure}
  \begin{center}\setlength{\piclength}{0.28\linewidth}
    \begin{tabular}{c@{\qquad}c@{\qquad}c}
      $(a)$ & $(b)$ & $(c)$ \\
      \includegraphics[width=\piclength]{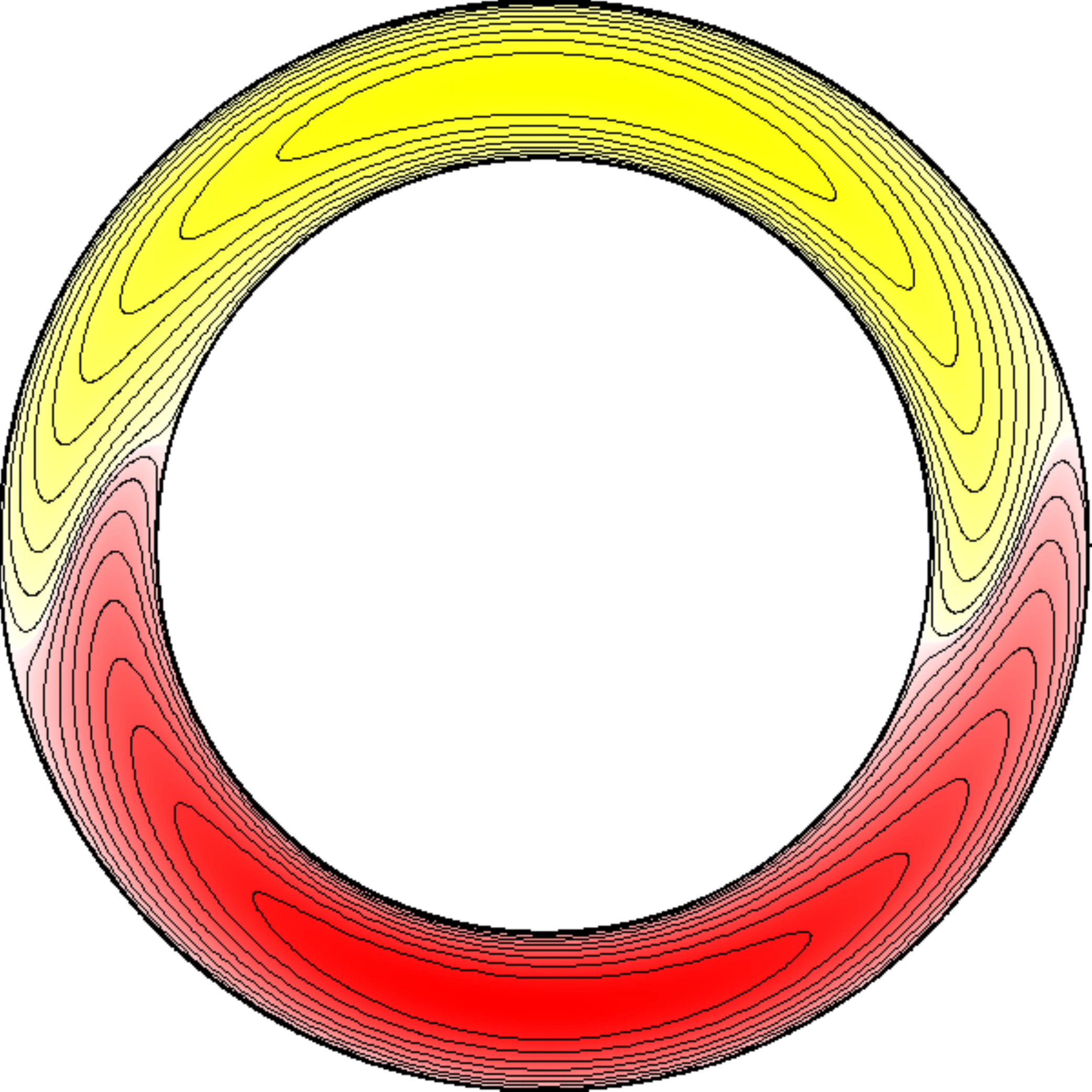} &
      \includegraphics[width=\piclength]{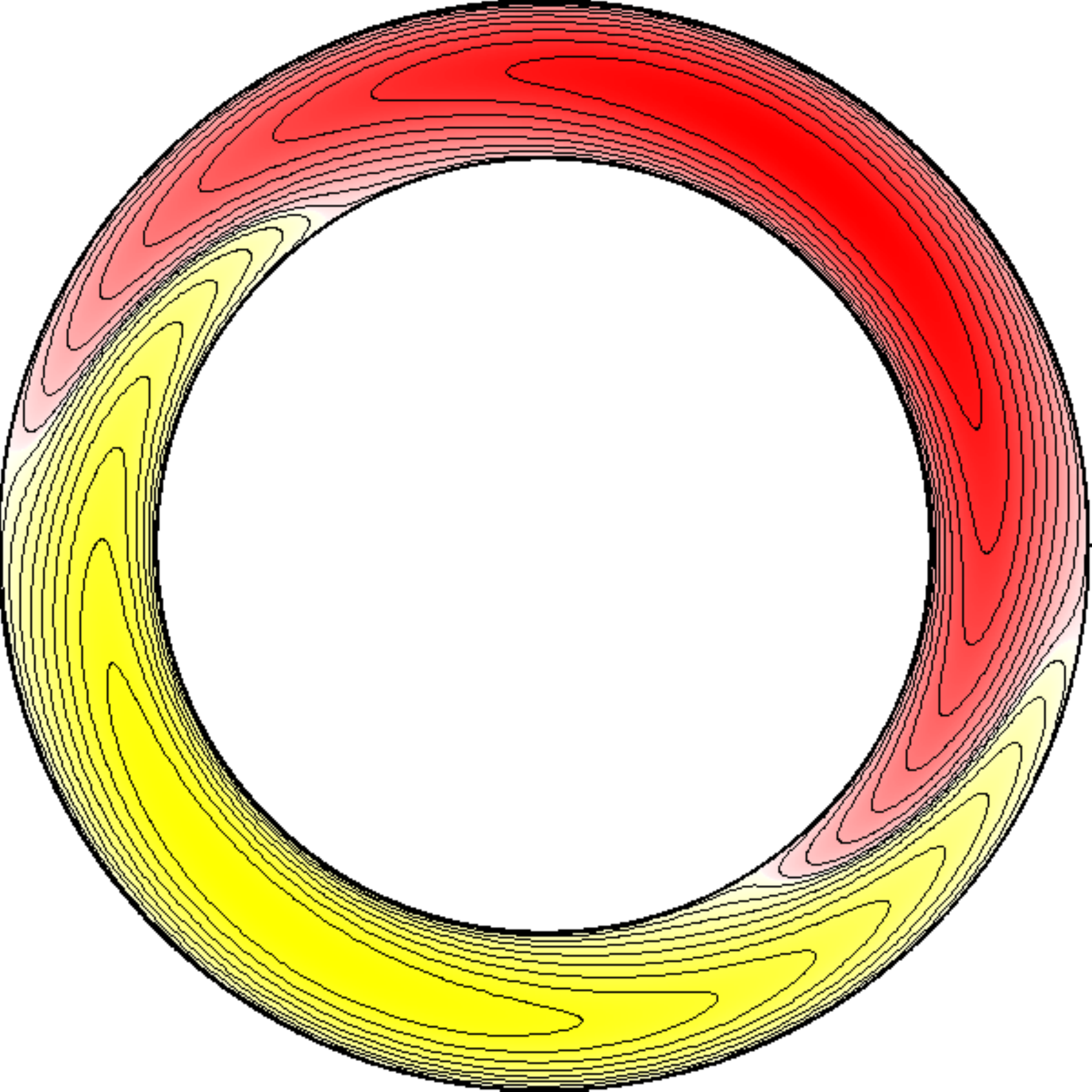} &
      \includegraphics[width=\piclength]{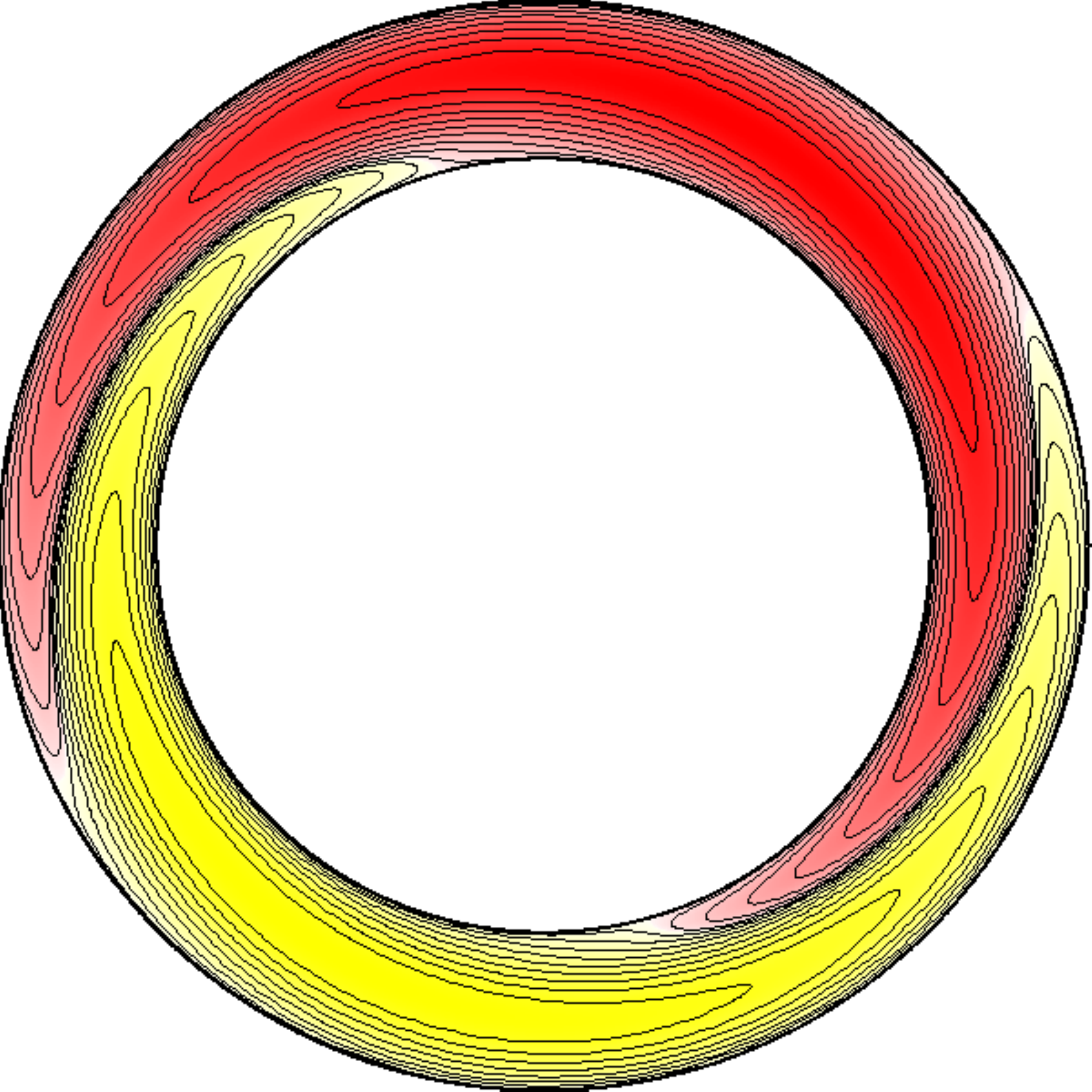} 
    \end{tabular}
  \end{center}
  \caption{Contours of the temperature disturbance $T'$ at a
    $z$-constant section corresponding to the points marked as blue
    circles in figure~\ref{Gr_Re_nosh}. $(a)$: $\Rey_i=5\times 10^5$,
    $G_c=21206.53$.  $(b)$: $\Rey_i=5.7\times 10^5$,
    $G_c=33768.37$. $(c)$: $\Rey_i=5.2\times 10^5$,
    $G_c=79670.16$. There are 10 positive (dark gray; red in the
    online version) and 10 negative (light gray; yellow in the online
    version) linearly spaced contours. In all cases the critical
    azimuthal mode is $n=1$ and $k=O(10^{-3})$.}
  \label{n1nosh}
\end{figure}

In this case a rotating frame of reference is readily identified and
the shear term $B/r$ in the base flow azimuthal velocity \eqref{ut} is
zero, whereas the term $A$ corresponds to the angular velocity of the
cylinders. Figure~\ref{Gr_Re_nosh} shows the critical values of $G$ as
the rotation speed, indicated here by the inner cylinder Reynolds
number $\Rey_i$, is increased. In the case of stationary cylinders
instability sets in at $G=8087.42$, with $k_c=-2.74$ and $n=0$. The
emerging pattern is characterized by pairs of counter-rotating
torodial rolls, that unlike Taylor vortices have a non-zero phase
velocity that causes a slow drift of the cellular pattern
upward. Extensive information about natural convection instabilities
can be found in the literature: \citet{ChKp80} and \citet{McCoBo84}
for infinite geometries, and \citet{VaTh69} and \citet{LKH82} for
finite geometries. Without rotation, traditional (dashed line) and new
(solid line) approximations yield identical results to the case where
centrifugal buoyancy is neglected (dashed-dotted line). For slow
rotation the effect of the centrifugal buoyancy is negligible, and
nearly the same critical values are obtained in each case (see inset
in figure~\ref{Gr_Re_nosh}). As rotation is increased, the flow is
strongly stabilized by centrifugal bouyancy. Note that if this is
neglected, the onset of instability asymptotically approaches
$G_c=172.50$ and is qualitatively wrong. The presence of the
centrifugal term in any of the ways considered here, entirely modifies
the stability of the problem and consequently is an essential element
to study these flows.  No differences between the two approximations
in the linear behaviour of the system are observed up to $\Rey_i\sim
5\times10^5$, where the two curves start to depart from each other.
Up to this point and after a small initial region where several
azimuthal modes up to $n=6$ are involved, the base flow loses
stability to an azimuthal mode $n=1$ with small axial wavenumber $k
\sim 10^{-3}$. The shape of the critical modes along the stability
curve is illustrated in figure~\ref{n1nosh}, showing contours of
constant temperature in a horizontal cross-section. The three states
correspond to the circles in figure~\ref{Gr_Re_nosh} and depict the
transition between the lower and intermediate branches as we consider
the new approximation.  As we proceed forward along the critical curve
the cold fluid progressively penetrates into the warm fluid and vice
versa. The same behaviour is observed when the traditional
approximation is used, nevertheless the values of $\Rey_i$ and $G_c$
required are larger.

As $\Rey_i$ increases beyond $5\times10^5$ the new terms in our
approximation start becoming important and lead to different behaviour
in the linear stability of the system. An analysis of the magnitude of
each term in our approximation reveals that the differences observed
in figure~\ref{Gr_Re_nosh} at high $\Rey_i$ are due to terms involving
the product $v_b u_\theta$, implying the existence of an important
centrifugal force acting in azimuthal direction as high rotational
speeds are reached. This provides evidence that the traditional
formulation, including only the main (radial) contribution of
centrifugal buoyancy, is a very good approximation if slow rotation is
involved but other contributions may not be neglected in rapidly
rotating fluids. Once the critical values given by both approximations
differ, we can identify two interesting regions in parameter
space. For $\Rey_i\in [5 \times 10^5,7.7 \times 10^5]$ the traditional
Boussinesq approach yields larger critical $G$ than our
approximation, whereas for $\Rey_i > 7.7 \times 10^5$ the upper branch of
the new approximation yields much lower critical values. Moreover, the
differences keep increasing as $\Rey_i$ grows. 

\begin{figure}
  \begin{center}
    \begin{tabular}{cc}
      $(a)$ & $(b)$ \\
      \includegraphics[scale=0.27]{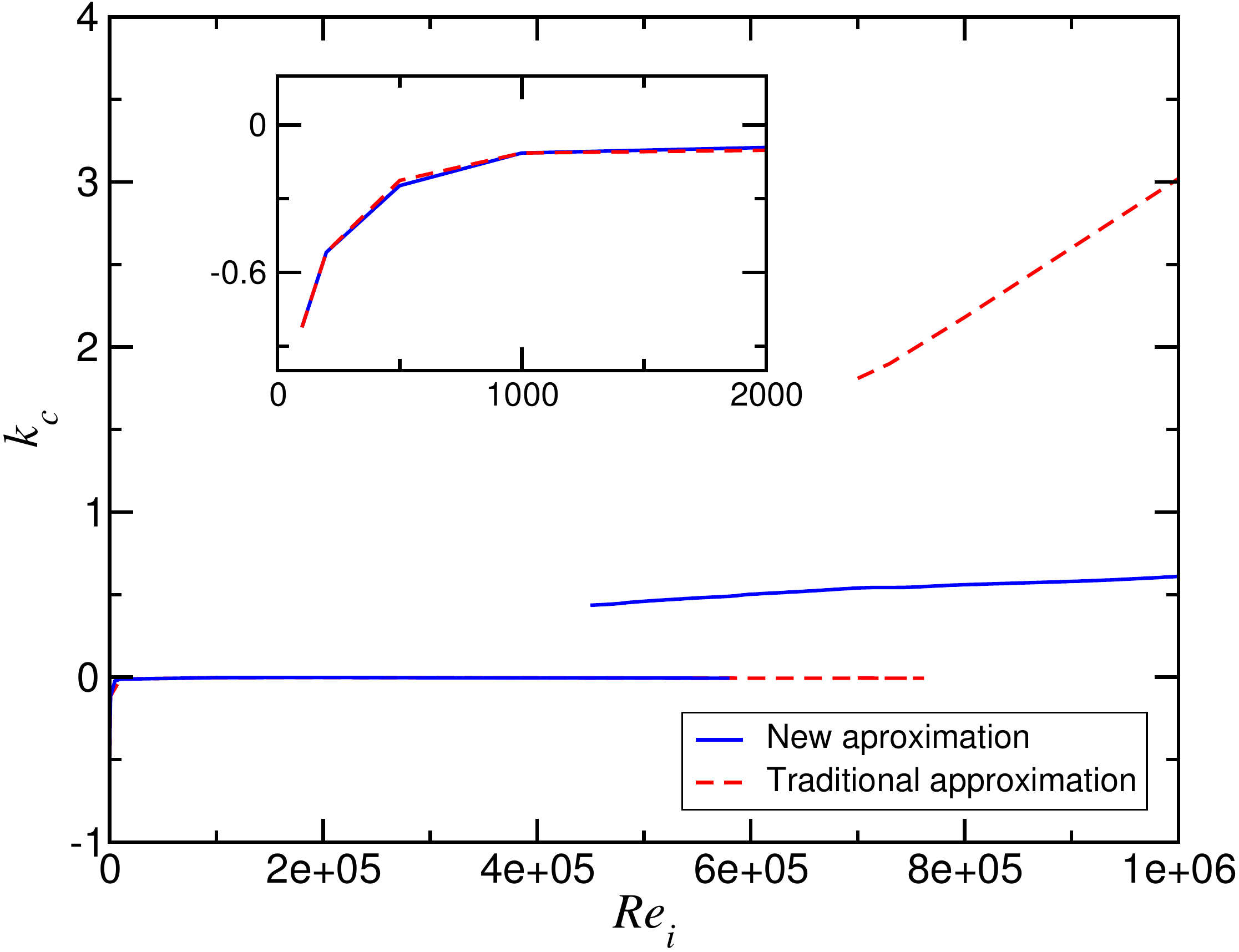} &
      \includegraphics[scale=0.27]{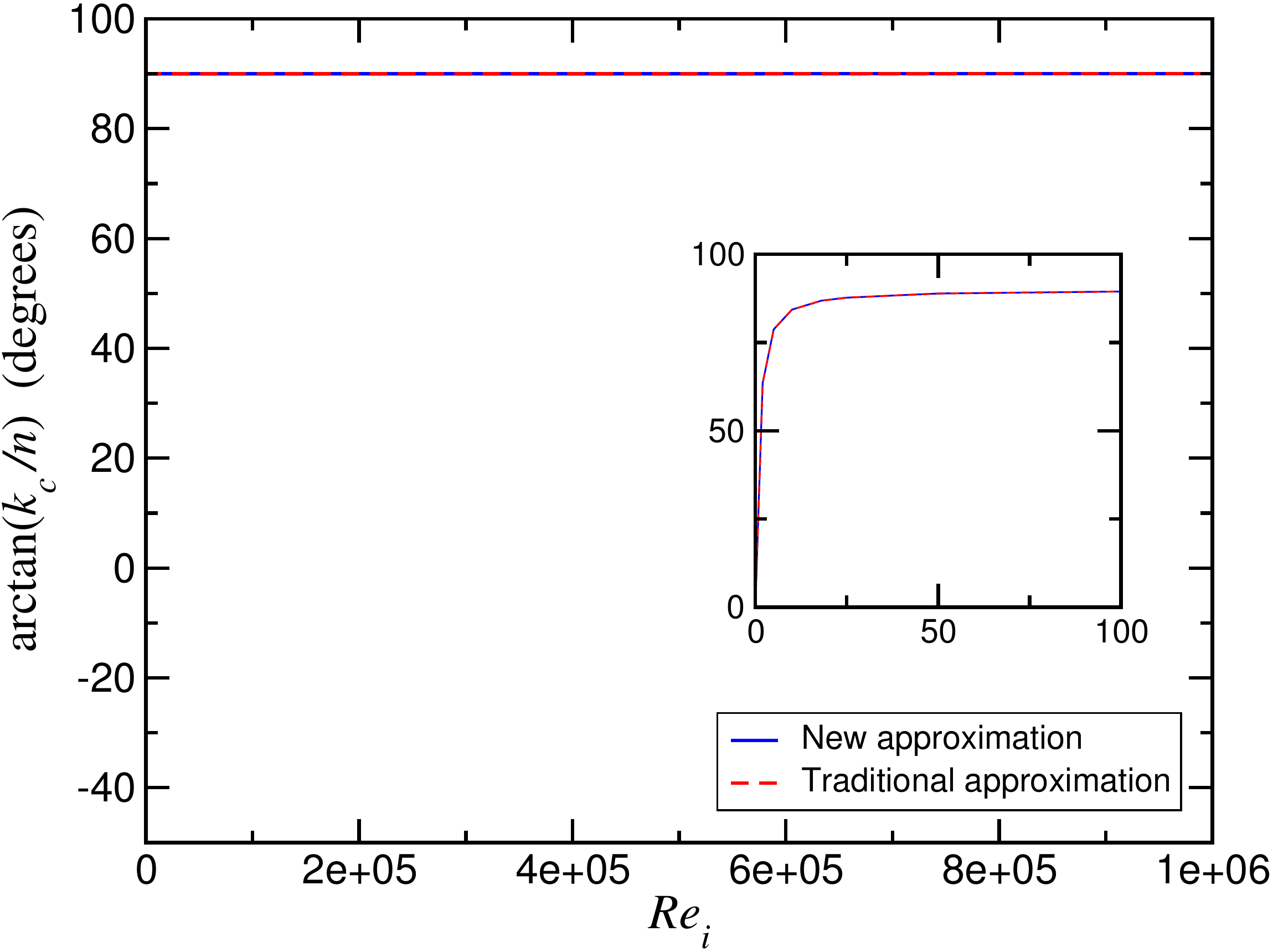} 
    \end{tabular}
  \end{center}
  \caption{$(a)$ Critical axial wavenumber $k_c$ and $(b)$ spiral
    angle of the modes $\arctan(k_c/n)$ as a function of $\Rey_i$ for
    the curves in figure~\ref{Gr_Re_nosh}. The inset is a close up at
    low $Re_i$ where the first mechanism stops being dominant and is
    superseded by spiral modes with angle far from $90$ degrees,
    indeed $0$, corresponding to $n=0$, at $Re_i=0$.}
  \label{kandmnosh}
\end{figure}

\begin{figure}
  \begin{center}
    \begin{tabular}{c@{\qquad}c}
      $(a)$ & $(b)$ \\
      \includegraphics[width=0.35\linewidth]{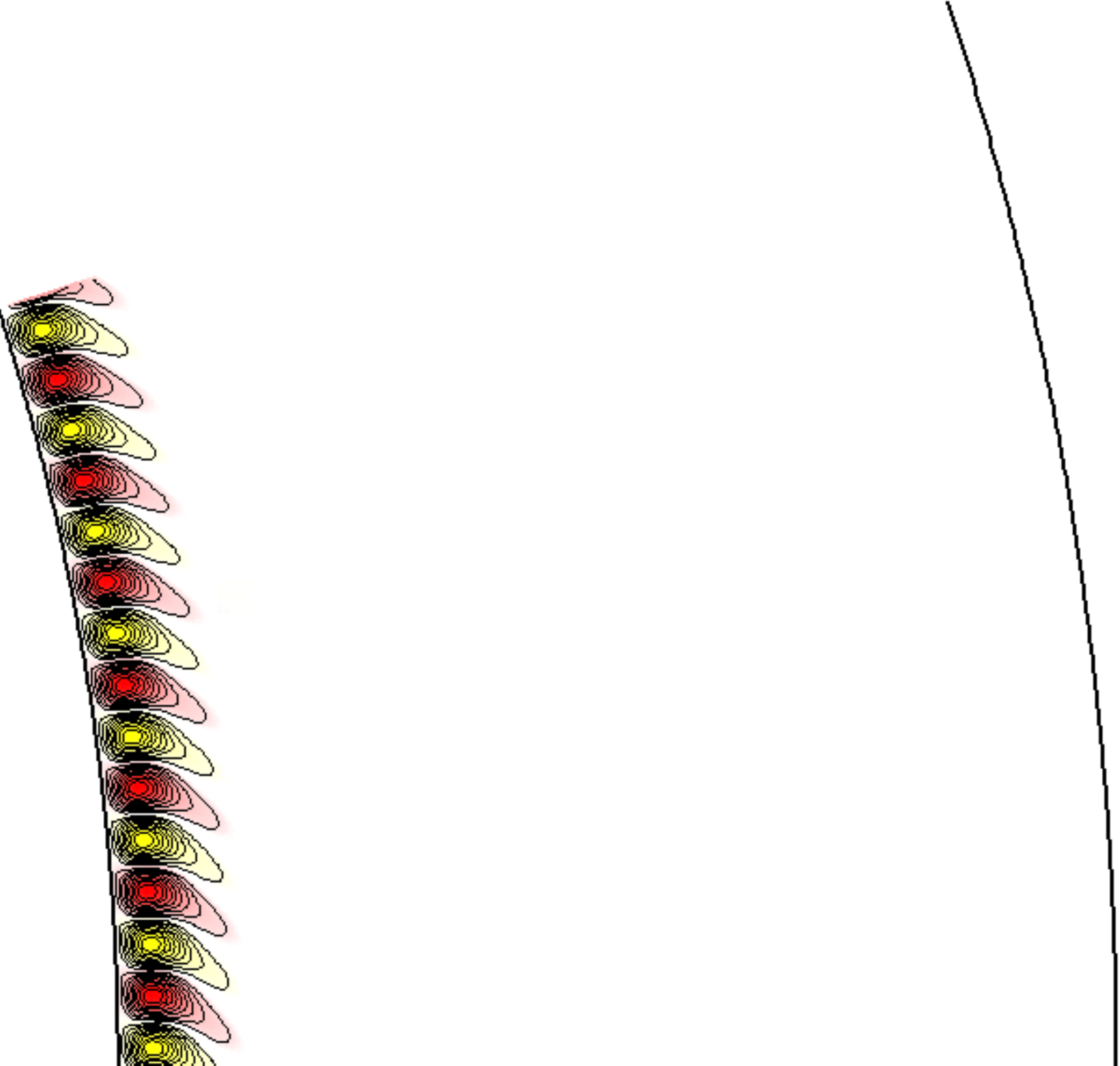} &
      \includegraphics[width=0.35\linewidth]{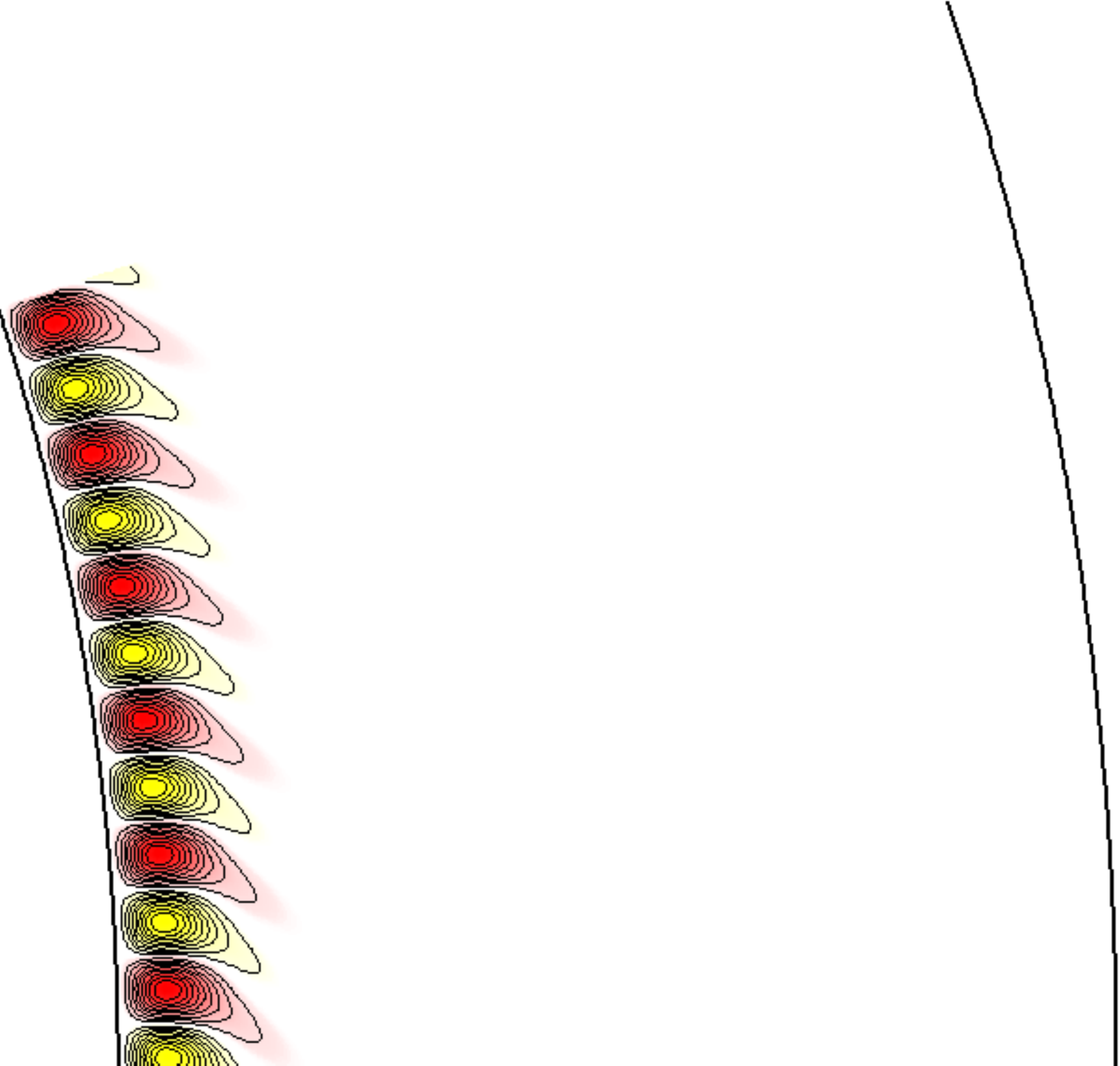} 
    \end{tabular}
  \end{center}
  \caption{Contours of the critical disturbance temperature on a
    $z$-constant section for $\Rey_i=700000$. $(a)$ Critical mode
    using the traditional Boussinesq approximation. Here
    $G_c=489371.47$, $k_c=1.81$ and $n=150$. $(b)$ Critical mode using
    the new approximation. Here $G_c=207906.92$, $k_c=0.54$ and
    $n=116$. In both cases, only $1/20$ of the domain is shown. There
    are 10 positive (darker gray; red online) and negative (light
    gray; yellow online) contours.}
  \label{highmodes}
\end{figure}

The analysis performed reveals the existence of two mechanisms of
instability associated with the lower-intermediate and upper branches
in figure~\ref{Gr_Re_nosh}. Different symbols are used to represent
the critical values corresponding to each mechanism in each
problem. The differences between them are illustrated in
figure~\ref{kandmnosh}, showing the evolution of the critical axial
wavenumber $k_c$ and the angle of the spiral modes
$\arctan(\frac{k_c}{n})$ versus $\Rey_i$. Two regions with distinct
characteristics are well-defined. The first mechanism of instability
has already been presented (see figure~\ref{n1nosh}). Low azimuthal
wavenumbers, primarily $n=1$, and very small axial wavenumbers
characterize it.  This corresponds to quasi two-dimensional modes and
can be readily seen in figure~\ref{kandmnosh}$(b)$, showing that the angle
of the spiral modes remains constant at about $90$ degrees. The inset
shows the small initial region where the spiral angle increases
progressively until it reaches a vertical position. The second
mechanism is characterized by $n>80$ and $k_c\sim O(1)$, also
corresponding to quasi two-dimensional modes (see
figure~\ref{kandmnosh}$b$). Another common feature between the two types
of instabilities is that the rotational frequency coincides with the
angular velocity of the container in both mechanisms and both
approximations. This is in agreement with \citet{MHA13}, who
have analytically proven that two-dimensional modes with $k=0$ always
rotate at speed $A$ \eqref{ut} in Taylor--Couette flows without
heating. An interesting distinct feature of the second instability
mechanism is localization near the inner cylinder. An example of these
wall convection modes is shown in figure~\ref{highmodes}; the critical
disturbances are clearly different in the traditional and in the new
Boussinesq approximations.

\subsection{Differentially rotating cylinders}

The traditional approximation for the centrifugal buoyancy neglects
the part of the base flow containing shear, \ie the $B/r$ term in
\eqref{ut}. To quantify the influence of including shear in the
centrifugal terms, we perform the same analysis as in the previous
section but for differentially rotating cylinders. The amount of shear
introduced is characterized by the ratio of angular velocities
$\beta=\Omega_i/\Omega_o$; the further $\beta$ is from one, the stronger
is the shear effect considered.

\subsubsection{Weak shear: rotation close to solid body ($\beta=\Omega_i/\Omega_o=1.006$)}

\begin{figure}
  \begin{center}
    \includegraphics[width=0.6\linewidth]{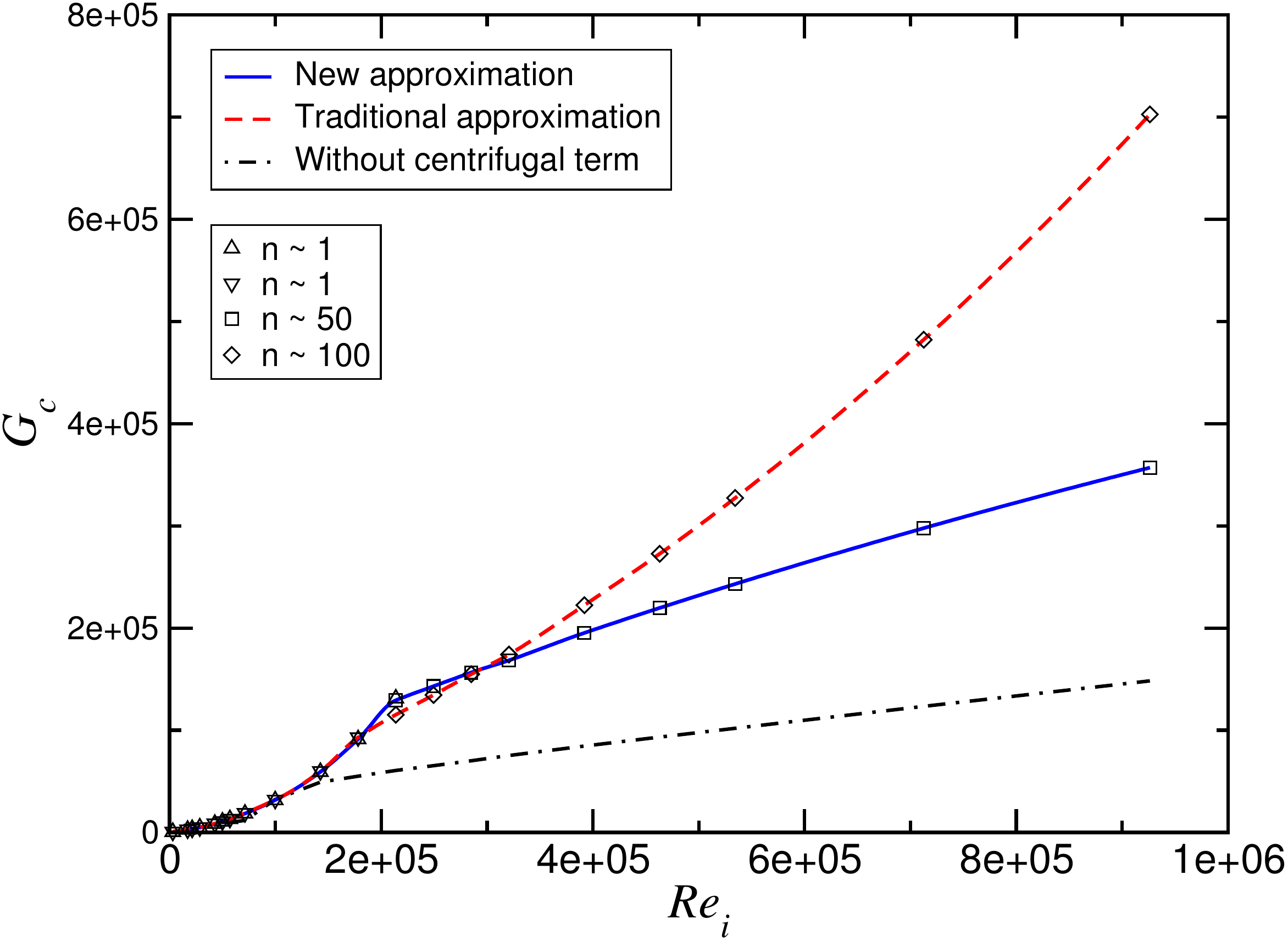}
  \end{center}
  \caption{Critical Grashof number $G_c$ as function of inner cylinder
    Reynolds number $Re_i$ for rotation near solid-body
    ($\beta=1.006$). Different symbols refer to two distinct instability
    mechanisms as in figure~\ref{Gr_Re_nosh}.}
  \label{Gr_Re_sh-116}
\end{figure}

We first consider the case where the container is rotating near solid
body. Although shear may be here expected to play only a secondary
role, this case serves the purpose of illustrating the importance of
including shear effects in the centrifugal term.
Figure~\ref{Gr_Re_sh-116} shows the neutral stability curve for the
two approaches considered and also without centrifugal buoyancy
(dashed-dotted line), which produces qualitatively correct results in
this instance. Unlike in the solid-body case, the critical values $G_c$
increase monotonically as $\Rey_i$ grows. Besides shear, centrifugal
effects are also important in this configuration. From $\Rey_i
\gtrsim 2\times10^5$ on the linear stability curves obtained by using
both approximations become quite different. Similar features with
respect to the solid-body case may be identified. At first the
traditional approximation gives lower critical value of the
Grashof. However, this region is smaller than in the solid-body case
and ends at $\Rey_i \sim 2.9\times10^5$ where both curves
intersect. From that point on, the stability region predicted by the
new approximation is smaller; the differences between the critical
values given by both approximations keep increasing as larger $\Rey_i$
are considered.  At the point where both curves first depart from each
other $\Rey_i$ has half the value of that of the solid-body
case. Consequently, the rotational speeds for which the new
approximation is necessary are significantly smaller in the presence
of weak differential rotation.

\begin{figure}
  \begin{center}
    \begin{tabular}{c@{\quad}c@{\quad}c}
      $(a)$ & $(b)$ & $(c)$ \\
      \includegraphics[width=0.3\linewidth]{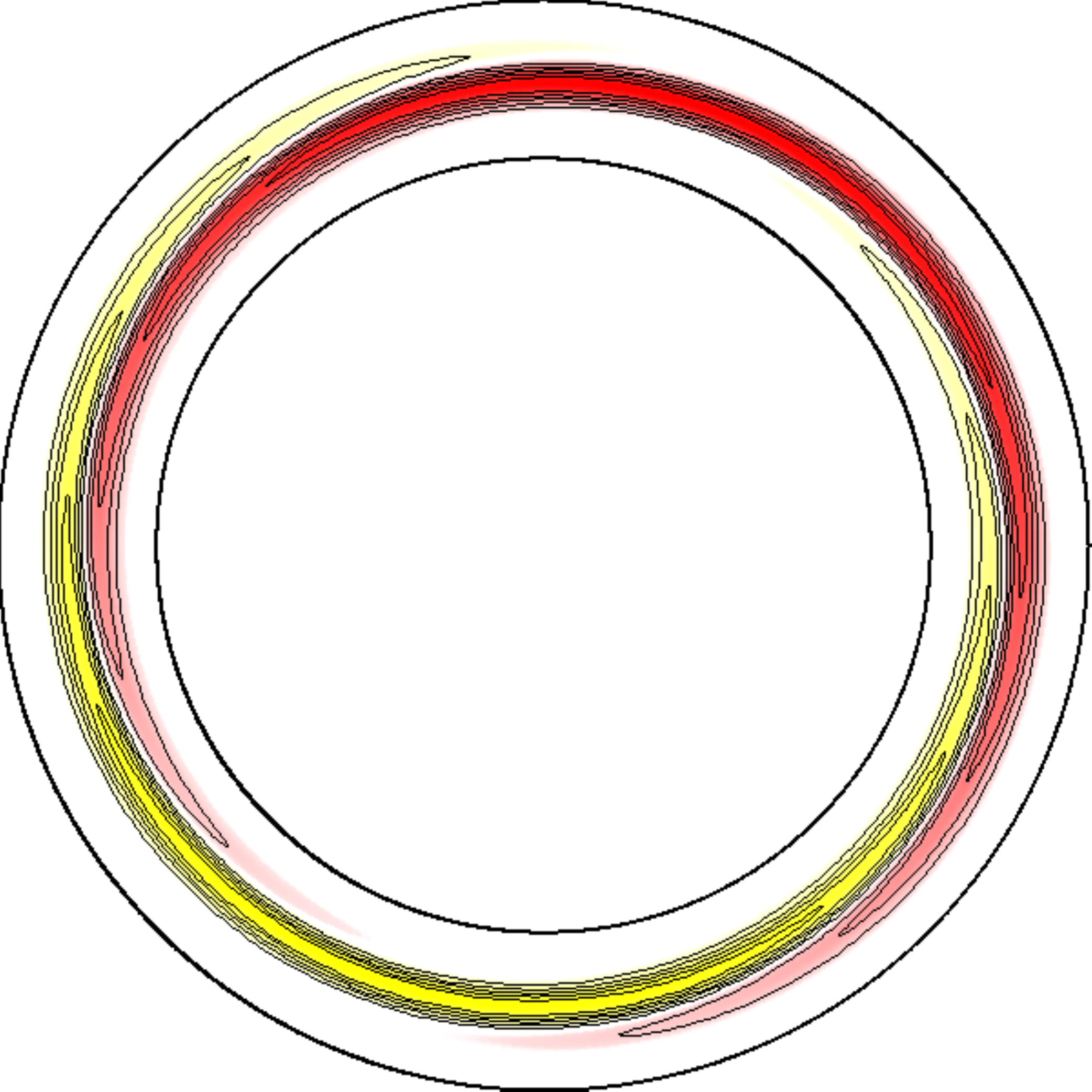} &
      \includegraphics[width=0.3\linewidth]{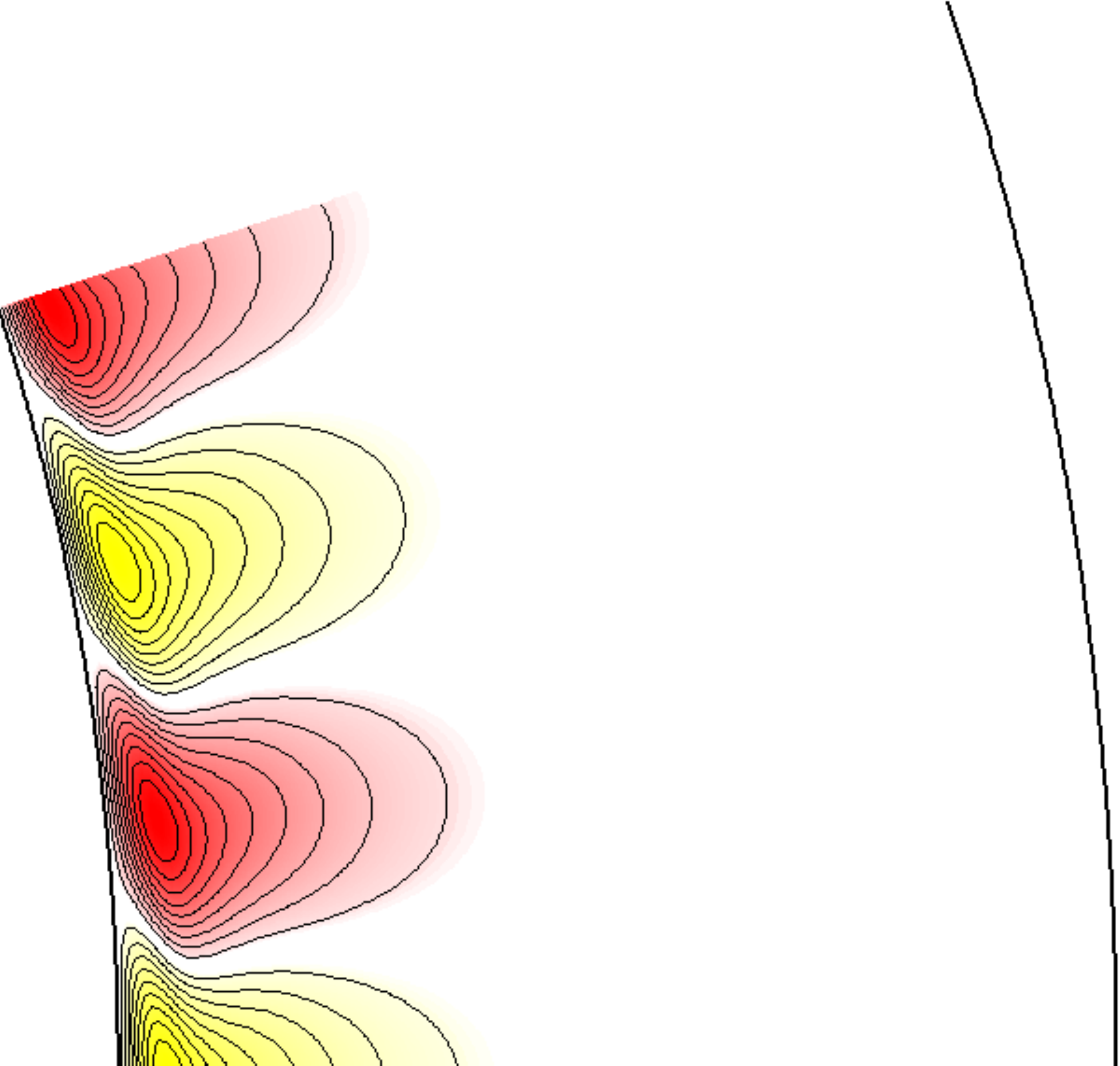} &
      \includegraphics[width=0.3\linewidth]{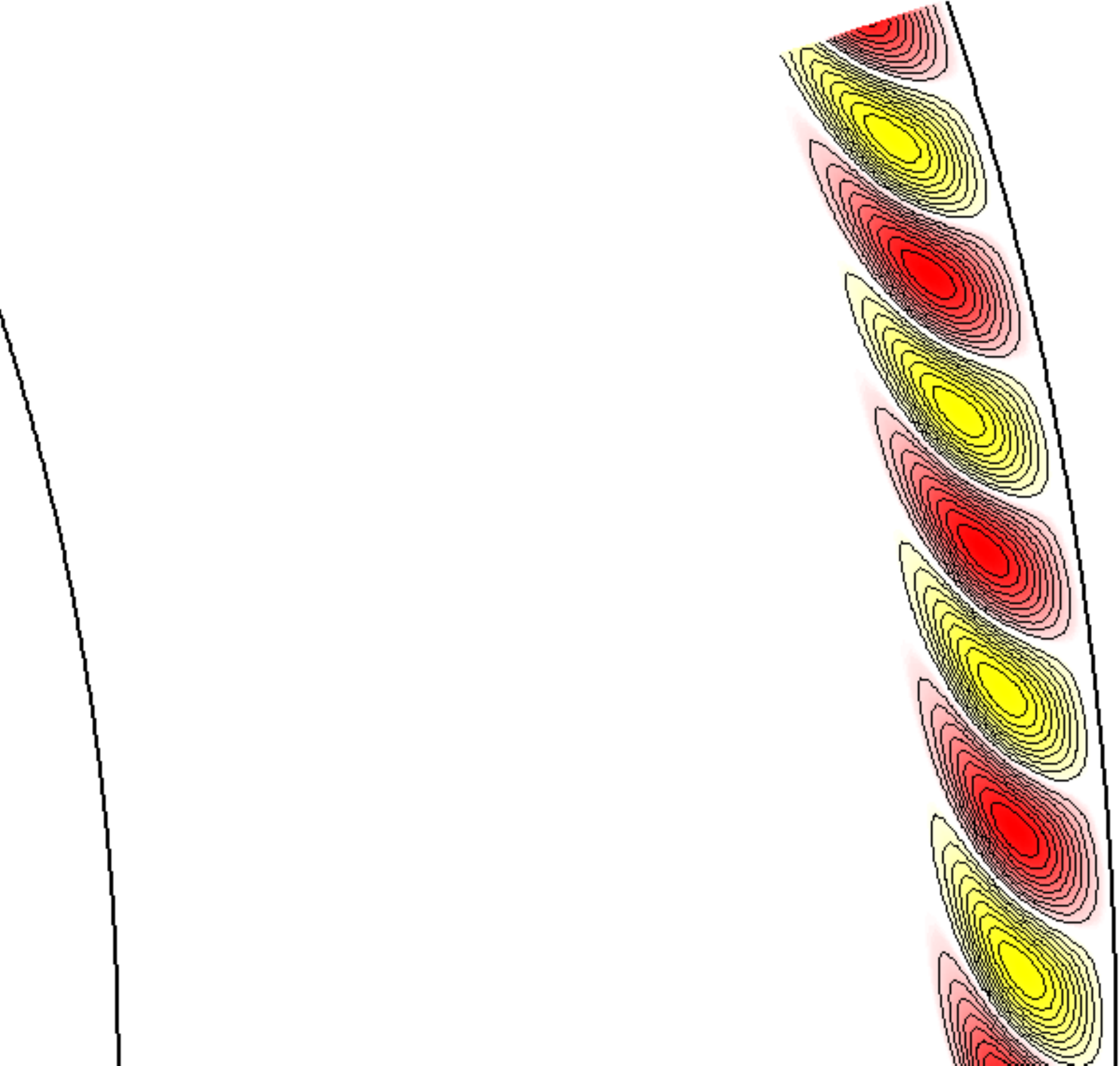} 
    \end{tabular}
  \end{center}
  \caption{Contours of the critical disturbance temperature on a
    $z$-constant section. $(a)$ $n=1$, $\Rey_i=178125$,
    $G_c=92987.56$. $(b)$--$(c)$ Comparison of the traditional $(b)$
    and new $(c)$ approximation at $\Rey_i=285000$ (near the crossover
    point in figure~\ref{Gr_Re_sh-116}) showing $1/20$ of the
    annulus. $(b)$ $G_c=154864.79$, $k_c=0.24$ and
    $n=30$. $(c)$ $G_c=156547.54$, $k_c=0.39$ and
    $n=75$. There are ten positive (dark gray; red online) and
    negative (light gray; yellow online) linearly spaced contours.}
  \label{sh-116flow}
\end{figure}

Critical axial and azimuthal wavenumbers exhibit similar behaviour to
the solid-body case and so they are not shown here. Two mechanisms of
instability are also found. The first one embraces the region
$2\times10^5 < \Rey_i$ and is characterized by $k_c \sim 0$ and $1
\leq n \leq 6$. Modes are similar to those obtained for the first
mechanism in the solid body case. A subtle difference can be
nevertheless pointed out. In the solid body situation the temperature
disturbances fill the whole annulus, whereas differential rotation
confines the perturbation towards the central part (see
figure~\ref{sh-116flow}$a$). The second mechanism also presents the
same features as in the solid body case, high azimuthal modes and
$k_c\in[0.5,1.5]$, but differences in the flow appear that deserve to
be highlighted. In the traditional approximation the dominant wall
modes are located at the inner cylinder, as it occurs in the
solid-body case (figure~\ref{sh-116flow}$b$). In contrast, using the
new approximation changes the location of the dominant wall modes to
the outer cylinder (figure~\ref{sh-116flow}$c$). In view of these
results we can say that considering shear effects in the centrifugal
term of the Navier--Stokes equations may be extremely important: not
only regarding the linear stability boundary but also the shape and
location of the critical modes.

\subsubsection{Strong shear: Quasi-Keplerian rotation ($\beta=\Omega_i/\Omega_o=1.58$)}

\begin{figure}
  \begin{center}
    \includegraphics[width=0.6\linewidth]{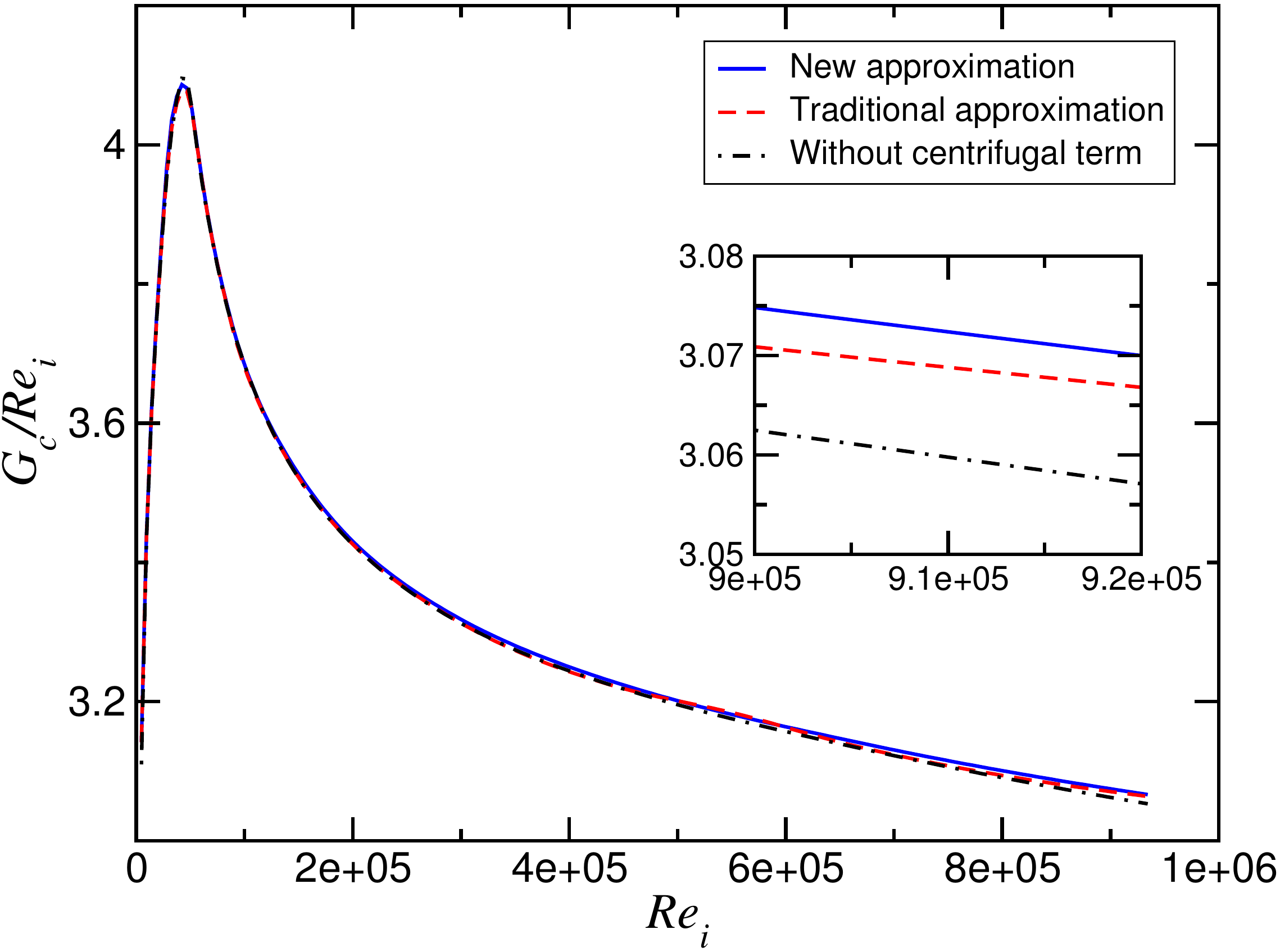}
  \end{center}
  \caption{Critical ratio $G_c/Re_i$ as function of inner cylinder
    Reynolds number $Re_i$ for quasi-Keplerian rotation
    ($\beta=1.58$). The three curves differ only by about 1\% and
    hence are only distinguishable in the inset.}
  \label{Gr_Re_sh-1.5}
\end{figure}

If $1/\eta > \beta > 1$ the angular velocity decreases outward but the
angular momentum increases. These flows, known as quasi-Keplerian
flows, are used as models to investigate the dynamics and stability of
astrophysical accretion disks. Here we choose a typical value
$\beta=1.58$ and as in the previous sections consider a negative
temperature gradient in the radial direction, as expected in accretion
disks. Figure~\ref{Gr_Re_sh-1.5} shows the neutral stability curve
for the two approximations considered, as well as entirely neglecting
centrifugal effects ($\epsilon=0$). The three curves are almost
straight lines, that completely overlap in a plot $(G_c,Re_i)$. In
order to see the small differences that appear at large $Re_i$, we
have plotted in this case $G_c/Re_i$ versus $Re_i$.  Shear is the
completely dominant mechanism in this regime, but small differences
can be observed for $Re_i\gtrsim 2\times10^5$, that are enhanced in
the inset. Surprisingly, shear has a very strong stabilizing effect in
this problem: without shear the critical Grashof number is ten times
smaller at $Re_i=10^6$ than in the quasi-Keplerian case.

\begin{figure}
  \begin{center}
    \begin{tabular}{c@{\quad}c}
      $(a)$ & $(b)$ \\
      \includegraphics[width=0.35\linewidth]{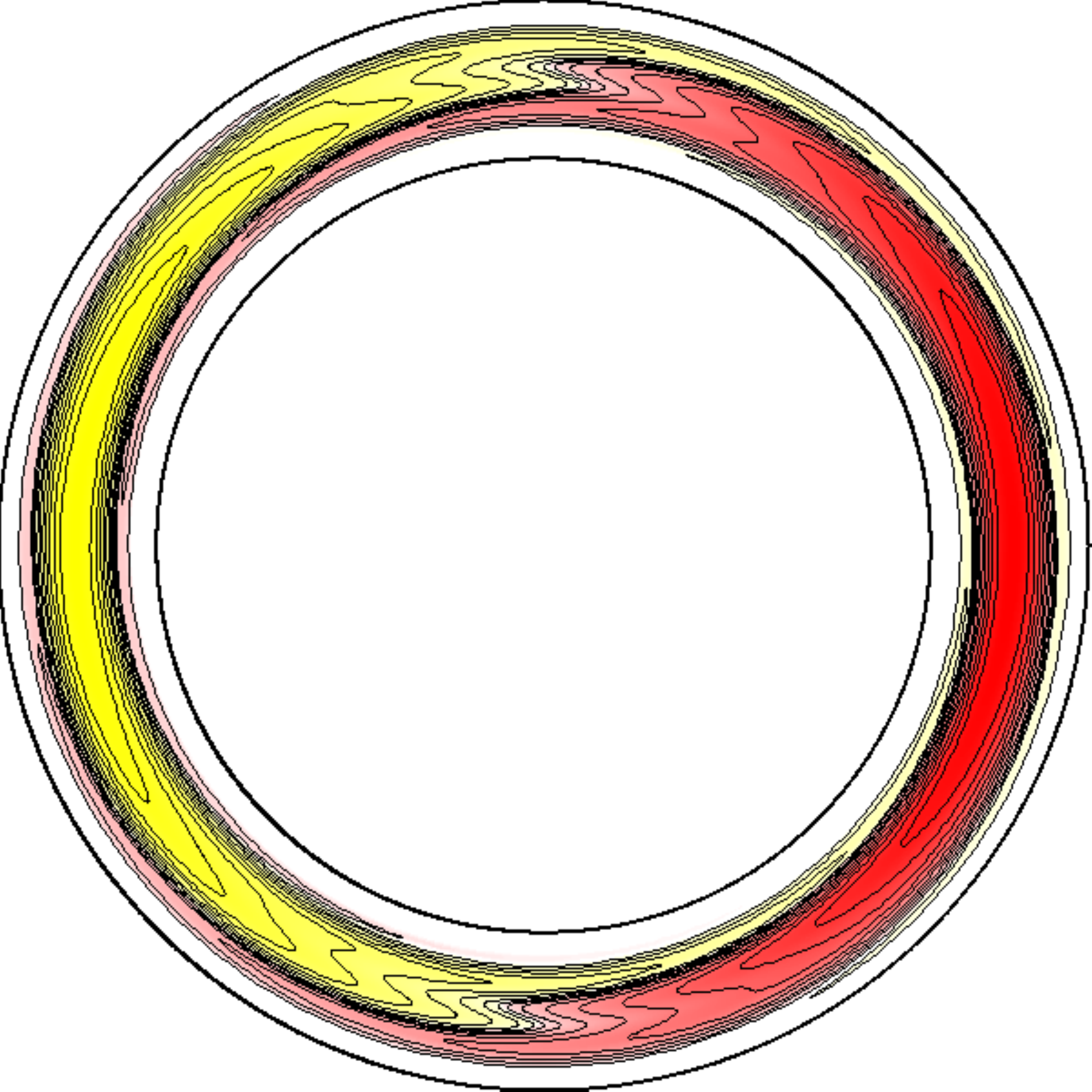} &
      \includegraphics[width=0.35\linewidth]{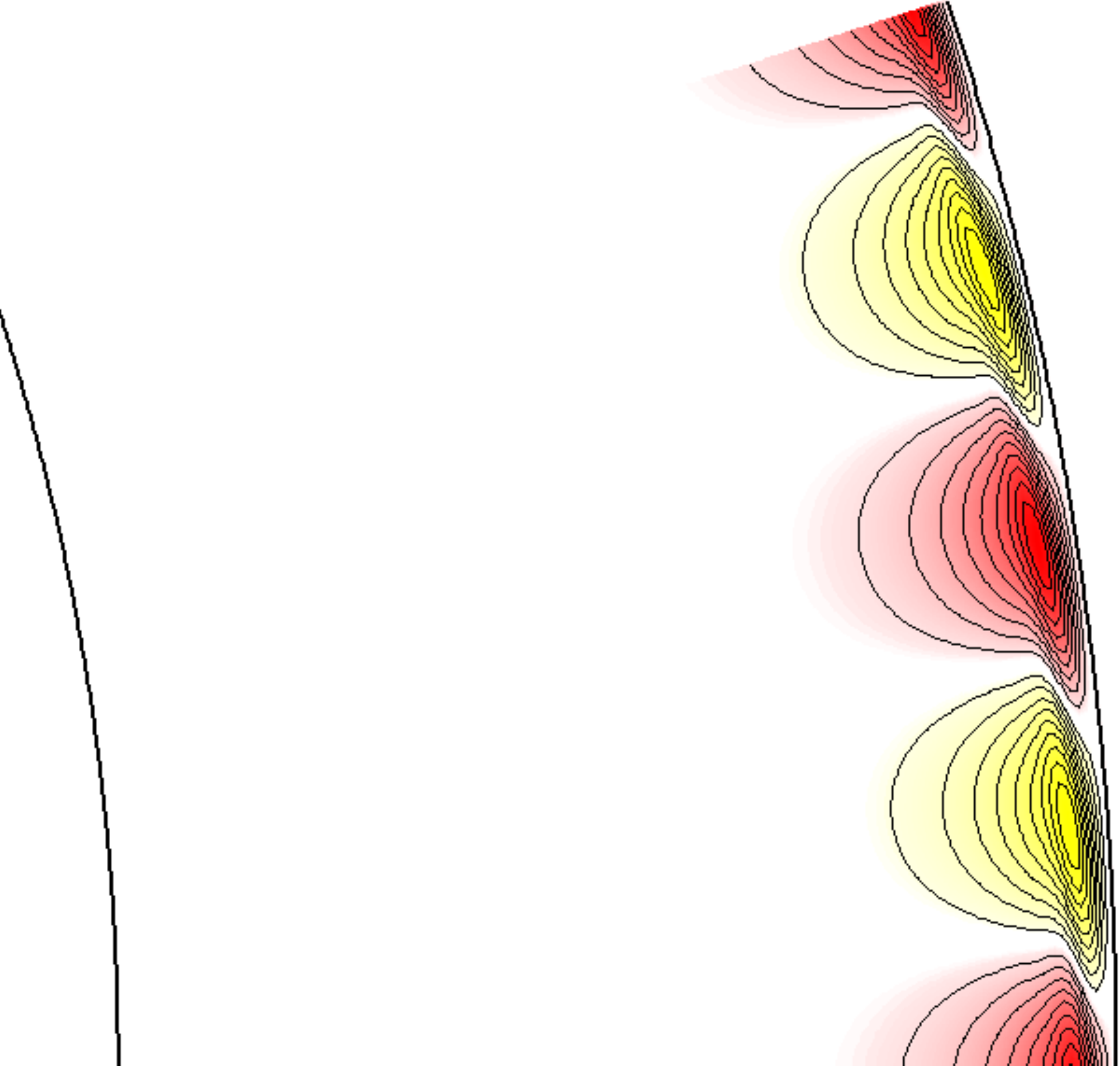} 
    \end{tabular}
  \end{center}
  \caption{Contours of the critical disturbance temperature on a
    $z$-constant section.  $(a)$ $\Rey_i=11681.03$ with
    $G_c=4.1268\times10^4$, $k_c=-1.05$ and $n=1$. $(b)$
    $\Rey_i=584051.72$ with $G_c=1.8511\times10^4$, $k_c=8.21$ and
    $n=38$. Ten positive (dark gray; red online) and negative contours
    (light gray; yellow online) are displayed. Only $1/20$ of the
    domain is shown in $(b)$.}
  \label{sh1p5flow}
\end{figure}

Depending on the Reynolds number two mechanisms of instability are
again found. The first mechanism exhibits a similar flow structure to
that observed in the previous case. It also occurs at low $\Rey_i$ and
is localized in the central part of the annulus due to the action of
differential rotation. Figure~\ref{sh1p5flow}$(a)$ shows the contours
of the disturbance temperature in an horizontal plane. In contrast to
what happens in the weak shear situation, these modes present a clear
3D structure with $k_c \sim -1$. Small azimuthal wavenumbers are
involved in this mechanism, ranging from $n=1$ to $n=6$. More
remarkable differences are found when analysing the second
mechanism. High azimuthal modes $n \sim 50$ arise as this mechanism
becomes dominant, but unlike the solid-body and weak shear situations,
the azimuthal mode number decreases as $\Rey_i$ increases.  The same
behaviour is observed in the axial wavenumber, so that the spiral
angle quickly converges to $90$ degrees as observed in the previous
sections. Figure~\ref{sh1p5flow}$(b)$ shows that the instability is
characterized by convection wall modes localized at the outer
cylinder, as in the case of weak shear using the new
approximation. Nevertheless, in quasi-Keplerian flows the dominant
modes are always localized at the outer cylinder regardless of how
centrifugal terms enter the equations.

%%%%%%%%%%%%%%%%%%%%%%%%%%%%%%%%%%%%%%%%%%%%%%%%%%%%%%%%%%%%%%%%%%%%%%

\section{Summary and discussion}

We have identified weaknesses in how the Boussinesq formulation is
typically used to account for centrifugal buoyancy in the
Navier--Stokes equations. In particular, the traditional approximation
(including only the term $\rho'\Omega^2$) neglects the effects
associated with differential rotation or strong internal
vorticity. This has motivated us to develop a new consistent
Boussinesq-type approximation correcting this problem. It consists in
keeping the density variations in the advection term of the Navier-Stokes
equations and thus it is very easy to implement in an existent
solver. The new approximation allows accurate treatment of situations
with differential rotation or when strong vortices appear in the
interior of the domain, which may cause important centrifugal effects
even in flows without global rotation. The latter may be especially
relevant in simulations at high Rayleigh numbers \citep[as e.g. in the
  quest for the `ultimate regime',][]{AGL09}. Thus we argue that our
formulation for the centrifugal terms should be always implemented
whenever the Boussinesq approximation is used.

The relevance of the new approximation has been illustrated with a
linear stability analysis of a Taylor--Couette system subjected to a
negative radial gradient of temperature. Three different cases have
been studied. First, we have considered the container rotating as
solid-body, \ie without differential rotation. We note that if
centrifugal buoyancy is entirely neglected, the results are even
qualitatively wrong. For both traditional and new approximations the
critical values obtained agree up to $\Rey_i \sim 5.5 \times 10^5$,
beyond which discrepancies become significant. Beyond this point the
conductive base flow loses stability to quasi two-dimensional wall
modes (aligned with the axis of rotation, as expected from the
Taylor--Proudman Theorem) localized at the inner cylinder. Note that
the large discrepancy in critical Grashof numbers observed at
$\Rey_i\in[5\times 10^5,10^6]$ between both approximations makes it
possible to test them against laboratory experiments. For example, in
the experiments from \citet{PaLa11}, and \cite{kavila2013}, which
allow for radial temperature gradients, $\Rey=10^6$ can be reached,
and the required Grashof numbers $5\times10^5$ can be obtained with
temperature differences about half a degree kelvin.

We have also considered the case in which the cylinders rotate at
different angular speeds, thus introducing shear. For weak
differential rotation, shear and centrifugal buoyancy effects compete
and the critical values obtained with both approximations differ from
each other at lower $\Rey_i\sim 2\times10^5$.  Moreover, the new
approximation gives rise to wall modes located on the outer cylinder,
whereas the traditional approach yields wall modes on the inner
cylinder, as in the solid-body case. In quasi-Keplerian flows, shear
is so dominant that centrifugal terms may be entirely neglected in the
linear stability analysis (discrepancies in $G_c$ are below 1\%
regardless of how centrifugal terms enter the equations, if at
all). Here the critical modes are always localized at the outer
cylinder. Note that such wall modes, similar to those identified by
\citet{klahr1999}, are not relevant to the accretion disk problem, in
which there are no solid radial boundaries. Furthermore, it is worth
noting that testing our differential rotation results in the
laboratory is very difficult because of axial endwall effects. The
large $\Rey$ involved will necessarily trigger instabilities and
transition to turbulence because of the nearly discontinuous angular
velocity profile at the junction between axial endwalls and cylinders
\citep{Avi12}.

Although it may be tempting to suggest that laminar quasi-Keplerian
flows are stable for weak stratification in the radial direction, our
analysis has only axial gravity, and is linear and hence concerned
with infinitesimal disturbances only. In more realistic models of
accretion disks, nonlinear baroclinic instabilities have been found in
similar regimes by \cite{KlBo03}, and we expect that subcritical
transition via finite amplitude disturbances may occur in the problem
investigated here. This remains a key question for incoming numerical
and experimental investigations. In fact, even in the classical
(isothermal) Taylor--Couette problem this possibility remains open and
controversial \cite[see e.g.][]{balbus2011}.

\begin{acknowledgements}
This work was supported by the Spanish Government grants FIS2009-08821
and BES-2010-041542. Part of the work was done during J.M.~Lopez's
visit to the Institute of Fluid Mechanics at the
Friedrich-Alexander-Universit\"at Erlangen-N\"urnberg, whose kind
hospitality is warmly appreciated.
\end{acknowledgements}   

%%%%%%%%%%%%%%%%%%%%%%%%%%%%%%%%%%%%%%%%%%%%%%%%%%%%%%%%%%%%%%%%%%%%%%


\begin{thebibliography}{35}
\expandafter\ifx\csname natexlab\endcsname\relax\def\natexlab#1{#1}\fi

\bibitem[Ahlers {\em et~al.\/}(2009)Ahlers, Grossmann \& Lohse]{AGL09}
{\sc Ahlers, Guenter, Grossmann, Siegfried \& Lohse, Detlef} 2009 Heat transfer
  and large scale dynamics in turbulent {R}ayleigh-{B\'e}nard convection. {\em
  Rev.\ Mod.\ Phys.\/} {\bf 81}~(2), 503--537.

\bibitem[Ali \& Weidman(1990)]{AlWe90}
{\sc Ali, M.~E. \& Weidman, P.~D.} 1990 On the stability of circular
  {C}ouette-flow with radial heating. {\em J.\,Fluid Mech.\/} {\bf 220},
  53--84.

\bibitem[Arfken \& Weber(2005)]{ArWe01}
{\sc Arfken, G.~B. \& Weber, H.~J.} 2005 {\em Mathematical Methods for
  Physicists\/}, 6th edn. Academic Press.

\bibitem[Avila \& Hof(2013)]{kavila2013}
{\sc Avila, Kerstin \& Hof, Bj{\"o}rn} 2013 {High-precision Taylor-Couette
  experiment to study subcritical transitions and the role of boundary
  conditions and size effects}. {\em Review of Scientific Instruments\/} {\bf
  84}, 065106.

\bibitem[Avila(2012)]{Avi12}
{\sc Avila, M.} 2012 Stability and angular-momentum transport of fluid flows
  between corotating cylinders. {\em Phys.\ Rev.\ Lett.\/} {\bf 108}, 124501.

\bibitem[Balbus(2003)]{Bal03}
{\sc Balbus, S.~A.} 2003 Enhanced angular momentum transport in accretion
  disks. {\em Annu.\ Rev.\ Astron.\ Astrophys.\/} {\bf 41}~(1), 555--597.

\bibitem[Balbus(2011)]{balbus2011}
{\sc Balbus, S.~A.} 2011 {Fluid dynamics: A turbulent matter}. {\em Nature\/}
  {\bf 470}~(7335), 475--476.

\bibitem[Bannon(1996)]{Ban96}
{\sc Bannon, PR} 1996 On the anelastic approximation for a compressible
  atmosphere. {\em J.\,Atmos.\ Sci.\/} {\bf 53}~(23), 3618--3628.

\bibitem[Barcilon \& Pedlosky(1967)]{BaPe67}
{\sc Barcilon, V. \& Pedlosky, J.} 1967 On the steady motions produced by a
  stable stratification in a rapidly rotating fluid. {\em J.\,Fluid Mech.\/}
  {\bf 29}, 673--690.

\bibitem[Batchelor(1967)]{Bat67}
{\sc Batchelor, G.~K.} 1967 {\em An introduction to Fluid Mechanics\/}.
  Cambridge University Press.

\bibitem[Boussinesq(1903)]{Bou03}
{\sc Boussinesq, J.} 1903 {\em Th\'eorie Analytique de la Chaleur\/}, ,
  vol.~II. Gauthier-Villars, Paris.

\bibitem[Brummell {\em et~al.\/}(2000)Brummell, Hart \& Lopez]{BHL00}
{\sc Brummell, N., Hart, J.~E. \& Lopez, J.~M.} 2000 On the flow induced by
  centrifugal buoyancy in a differentially-heated rotating cylinder. {\em
  Theoret.\ Comput.\ Fluid Dynamics\/} {\bf 14}, 39--54.

\bibitem[Canuto {\em et~al.\/}(2007)Canuto, Quarteroni, Hussaini \&
  Zang]{CQHZ07}
{\sc Canuto, C., Quarteroni, A., Hussaini, M.~Y. \& Zang, T.~A.} 2007 {\em
  Spectral Methods. Evolution to Complex Geometries and Applications to Fluid
  Dynamics\/}. Springer-Verlag.

\bibitem[Chandrasekhar(1961)]{Ch61}
{\sc Chandrasekhar, S.} 1961 {\em Hydrodynamic and Hydromagnetic Stability\/}.
  Oxford University Press.

\bibitem[Choi \& Korpela(1980)]{ChKp80}
{\sc Choi, I.G. \& Korpela, S.A.} 1980 Stability of the conduction regime of
  natural convection in a tall vertical annulus. {\em J.\,Fluid Mech.\/} {\bf
  99}, 725--738.

\bibitem[Elperin {\em et~al.\/}(1998)Elperin, Kleeorin \& Rogachevskii]{ElRo98}
{\sc Elperin, T., Kleeorin, N. \& Rogachevskii, I.} 1998 Dynamics of particles
  advected by fast rotating turbulent fluid flow: Fluctuations and large-scale
  structures. {\em Phys.\ Rev.\ Lett.\/} {\bf 81}, 2898--2901.

\bibitem[Hart(2000)]{Har00}
{\sc Hart, J.~E.} 2000 On the influence of centrifugal buoyancy on rotating
  convection. {\em J.\,Fluid Mech.\/} {\bf 403}, 133--151.

\bibitem[Hide \& Fowlis(1965)]{hide1965}
{\sc Hide, R. \& Fowlis, W.W.} 1965 {Thermal convection in a rotating annulus
  of liquid: Effect of viscosity on the transition between axisymmetric and
  non-axisymmetric flow regimes}. {\em Journal of Atmospheric Sciences\/} {\bf
  22}, 541--558.

\bibitem[Homsy \& Hudson(1969)]{HoHu69}
{\sc Homsy, G.~M. \& Hudson, J.~L.} 1969 Centrifugally driven thermal
  convection in a rotating cylinder. {\em J.\,Fluid Mech.\/} {\bf 35}, 33--52.

\bibitem[Klahr \& Bodenheimer(2003)]{KlBo03}
{\sc Klahr, H. \& Bodenheimer, P.} 2003 Turbulence in accretion disks:
  Vorticity generation and angular momentum transport via the global baroclinic
  instability. {\em The Astrophysical Journal\/} {\bf 582}~(2), 869.

\bibitem[Klahr {\em et~al.\/}(1999)Klahr, Henning \& Kley]{klahr1999}
{\sc Klahr, H.H., Henning, Th. \& Kley, W.} 1999 On the azimuthal structure of
  thermal convection in circumstellar disks. {\em The Astrophysical Journal\/}
  {\bf 514}~(1), 325.

\bibitem[Lee {\em et~al.\/}(1982)Lee, Korpela \& Horn]{LKH82}
{\sc Lee, Y., Korpela, S.~A. \& Horn, R.~N.} 1982 Structure of multicellular
  natural convection in a tall vertical annulus. In {\em Proc. 7th Zntl Heat
  Transfer Conf. Munich\/}, , vol.~2, pp. 221--226.

\bibitem[Lesur \& Papaloizou(2010)]{LePa10}
{\sc Lesur, G. \& Papaloizou, J. C.~B.} 2010 The subcritical baroclinic
  instability in local accretion disc models. {\em AA\/} {\bf 513}, A60.

\bibitem[Lopez \& Marques(2009)]{LoMa09}
{\sc Lopez, J.~M. \& Marques, F.} 2009 Centrifugal effects in rotating
  convection: nonlinear dynamics. {\em J.\,Fluid Mech.\/} {\bf 628}, 269--297.

\bibitem[Maretzke {\em et~al.\/}(2013)Maretzke, Hof \& Avila]{MHA13}
{\sc Maretzke, S., Hof, B. \& Avila, M.} 2013 Transient growth in linearly
  stable taylor--couette flows. {\em J.\,Fluid Mech.\/} p. submitted.

\bibitem[Marques {\em et~al.\/}(2007)Marques, Mercader, Batiste \&
  Lopez]{MMBL07}
{\sc Marques, F., Mercader, I., Batiste, O. \& Lopez, J.~M.} 2007 Centrifugal
  effects in rotating convection: Axisymmetric states and 3d instabilities.
  {\em J.\,Fluid Mech.\/} {\bf 580}, 303--318.

\bibitem[McFadden {\em et~al.\/}(1984)McFadden, Coriell, Boisvert \&
  Glicksman]{McCoBo84}
{\sc McFadden, G.~B., Coriell, S.~R., Boisvert, R.~F. \& Glicksman, M.~E.} 1984
  Asymmetric instabilities in buoyancy-driven flow in a tall vertical annulus.
  {\em Phys.\ Fluids\/} {\bf 27}, 1359--1361.

\bibitem[Meseguer {\em et~al.\/}(2007)Meseguer, Avila, Mellibovsky \&
  Marques]{MAMM07}
{\sc Meseguer, A., Avila, M., Mellibovsky, F. \& Marques, F.} 2007 {Solenoidal
  spectral formulations for the computation of secondary flows in cylindrical
  and annular geometries}. {\em European Physics Journal Special Topics\/} {\bf
  146}, 249--259.

\bibitem[Meseguer \& Marques(2000)]{MeMa00}
{\sc Meseguer, A. \& Marques, F.} 2000 On the competition between centrifugal
  and shear instability in spiral couette flow. {\em J.\,Fluid Mech.\/} {\bf
  402}, 33--56.

\bibitem[Paoletti \& Lathrop(2011)]{PaLa11}
{\sc Paoletti, M.~S. \& Lathrop, D.~P.} 2011 Measurement of angular momentum
  transport in turbulent flow between independently rotating cylinders. {\em
  Phys.\ Rev.\ Lett.\/} {\bf 106}, 024501.

\bibitem[Petersen {\em et~al.\/}(2007)Petersen, Julien \& Stewart]{PJS07}
{\sc Petersen, M., Julien, K. \& Stewart, G.} 2007 Baroclinic vorticity
  production in protoplanetary disks. {\em The Astrophysical Journal\/} {\bf
  658}, 1236--1251.

\bibitem[Randriamampianina {\em et~al.\/}(2006)Randriamampianina, Fr\"uh, Read
  \& Maubert]{RFRM06}
{\sc Randriamampianina, A., Fr\"uh, W.-G., Read, P.~L. \& Maubert, P.} 2006
  Direct numerical simulations of bifurcations in an air-filled rotating
  baroclinic annulus. {\em J.\,Fluid Mech.\/} {\bf 561}, 359--389.

\bibitem[Regev \& Umurhan(2008)]{ReUm08}
{\sc Regev, O. \& Umurhan, O.~M.} 2008 On the viability of the shearing box
  approximation for numerical studies of mhd turbulence in accretion disks.
  {\em Astron.\ \& Astrophys.\/} {\bf 481}~(1), 21--32.

\bibitem[Tassoul(2000)]{Tass00}
{\sc Tassoul, J.~L.} 2000 {\em Stellar Rotation\/}. Cambridge Univ. Press.

\bibitem[de~Vahl~Davis \& Thomas(1969)]{VaTh69}
{\sc de~Vahl~Davis, G. \& Thomas, R.~W.} 1969 Natural convection between
  concentric vertical cylinders. {\em Phys.\ Fluids Suppl.\ II\/} pp. 198--207.

\end{thebibliography}
\end{document}